\documentclass[acmsmall]{acmart}
\AtBeginDocument{%
  \providecommand\BibTeX{{%
    Bib\TeX}}}

\setcopyright{acmlicensed}
\copyrightyear{2025}
\acmYear{2025}
\acmDOI{XXXXXXX.XXXXXXX}

\acmJournal{JACM}
\acmVolume{37}
\acmNumber{4}
\acmArticle{111}
\acmMonth{9}

\usepackage{amsmath,amssymb,amsfonts}
\usepackage{algorithmic}
\usepackage{graphicx}
\usepackage{subcaption}
\usepackage{textcomp}
\usepackage{caption}
\usepackage{xcolor}
\usepackage{colortbl}
\usepackage{enumitem}
\usepackage{multirow}
\usepackage{url}
\usepackage{hyperref}
\usepackage{balance}
\usepackage[breakable]{tcolorbox}
\def\BibTeX{{\rm B\kern-.05em{\sc i\kern-.025em b}\kern-.08em
    T\kern-.1667em\lower.7ex\hbox{E}\kern-.125emX}}
\usepackage{xspace}
\newcommand{\mypara}[1]{\smallskip \noindent\textbf{#1.} \xspace}
\usepackage{longfbox}
\definecolor{ashgrey}{rgb}{0.7, 0.75, 0.71}
\definecolor{grey}{rgb}{0.6,0.6,0.6}

\definecolor{green1}{RGB}{200, 230, 200} % 浅绿色
\definecolor{green2}{RGB}{150, 210, 150} % 中等绿色
\definecolor{green3}{RGB}{100, 190, 100} % 较深的绿色
\definecolor{green4}{RGB}{50, 160, 50}
\definecolor{green5}{RGB}{240, 250, 240} % 非常浅的绿色

\begin{document}

% \title{An Experimental Evaluation of Ensembling Large Language Models for Source Code Vulnerability Detection}
\title{Ensembling Large Language Models for Code Vulnerability Detection: An Empirical Evaluation}

% \affiliation{%

\author{Zhihong Sun}
\affiliation{%
  \institution{Shandong Normal University}
  \city{Jinan}
  \country{China}}
\email{2022021002@stu.sdnu.edu.cn}

\author{Jia Li}
\affiliation{%
  \institution{College of AI, Tsinghua University}
  \city{Beijing}
  \country{China}}
\email{jia\_li@mail.tsinghua.edu.cn}

\author{Yao Wan}
\affiliation{%
  \institution{Huazhong University of Science and Technology}
  \city{Wuhan}
  \country{China}}
\email{wanyao@hust.edu.cn}

\author{Chuanyi Li}
\affiliation{%
  \institution{National Key Laboratory for Novel
Software Technology, Nanjing University}
  \city{Nanjing}
  \country{China}}
\email{lcy@nju.edu.cn;}

\author{Hongyu Zhang}
\affiliation{%
  \institution{Chongqing University}
  \city{Chongqing}
  \country{China}}
\email{hyzhang@cqu.edu.cn}

\author{Zhi Jin}
\affiliation{%
  \institution{Peking University}
  \city{Beijing}
  \country{China}}
\email{zhijin@pku.edu.cn}

\author{Ge Li}
\affiliation{%
  \institution{Peking University}
  \city{Beijing}
  \country{China}}
\email{lige@pku.edu.cn}

\author{Hong Liu}
\affiliation{%
  \institution{Shandong Normal University}
  \city{Jinan}
  \country{China}}
\email{lhsdcn@126.com}

\author{Chen Lyu}
\authornote{Chen Lyu is the corresponding author.}
\affiliation{%
  \institution{Shandong Normal University}
  \city{Jinan}
  \country{China}}
\email{lvchen@sdnu.edu.cn}

\author{Songlin Hu}
\affiliation{%
  \institution{Institute of Information Engineering, Chinese Academy of Sciences}
  \city{Beijing}
  \country{China}}
\email{husonglin@iie.ac.cn}

\renewcommand{\shortauthors}{Sun et al.}

\begin{abstract}
Code vulnerability detection is crucial for ensuring the security and reliability of modern software systems. Recently, Large Language Models (LLMs) have shown promising capabilities in this domain. However, notable discrepancies in detection results often arise when analyzing identical code segments across different training stages of the same model or among architecturally distinct LLMs. While such inconsistencies may compromise detection stability, they also highlight a key opportunity: the latent complementarity among models can be harnessed through ensemble learning to create more robust vulnerability detection systems. In this study, we explore the potential of ensemble learning to enhance the performance of LLMs in source code vulnerability detection. We conduct comprehensive experiments involving five LLMs (i.e., DeepSeek-Coder-6.7B, CodeLlama-7B, CodeLlama-13B, CodeQwen1.5-7B, and StarCoder2-15B), using three ensemble strategies (i.e., Bagging, Boosting, and Stacking). These experiments are carried out across three widely adopted datasets (i.e., Devign, ReVeal, and BigVul). Inspired by Mixture of Experts (MoE) techniques, we further propose Dynamic Gated Stacking (DGS), a Stacking variant tailored for vulnerability detection. Our results demonstrate that ensemble approaches can significantly improve detection performance, with Boosting excelling in scenarios involving imbalanced datasets. Moreover, DGS consistently outperforms traditional Stacking, particularly in handling class imbalance and multi-class classification tasks. These findings offer valuable insights into building more reliable and effective LLM-based vulnerability detection systems through ensemble learning.
\end{abstract}




% \received{20 February 2025}

\maketitle

\section{Introduction}

Code vulnerability detection is typically formulated as a classification task that determines whether a given function contains security vulnerabilities. By automatically identifying potentially exploitable vulnerable functions, this task is essential for enhancing the robustness, reliability, and trustworthiness of modern software systems~\cite{linevul, hanif2022vulberta, nguyen2022regvd}. 
In literature, numerous approaches have been proposed for vulnerability detection, utilizing both static analysis and machine learning techniques. 
% Traditional static analysis methods~\cite{brito2023study, kaur2020comparative, afrose2022evaluation} depend on human experts to manually create specific rules for identifying vulnerabilities. 
Traditional static analysis methods~\cite{brito2023study, kaur2020comparative, afrose2022evaluation} rely heavily on static analysis tools (e.g., pattern matching, syntax analysis) and human-crafted rules to identify vulnerabilities. However, these approaches are constrained by the limited scope of detectable vulnerabilities, as they often fail to recognize complex or novel vulnerability patterns beyond predefined rules. To this end, machine-learning-based approaches such as VulDeePecker~\cite{li2018vuldeepecker} and VulDeeLocator~\cite{li2021vuldeelocator}, have made significant advances, primarily attributed to their ability to learn comprehensive code representations, thereby enhancing detection capabilities across diverse types of vulnerability.

\begin{figure*}[htbp]
    \centering
    \includegraphics[width=\linewidth]{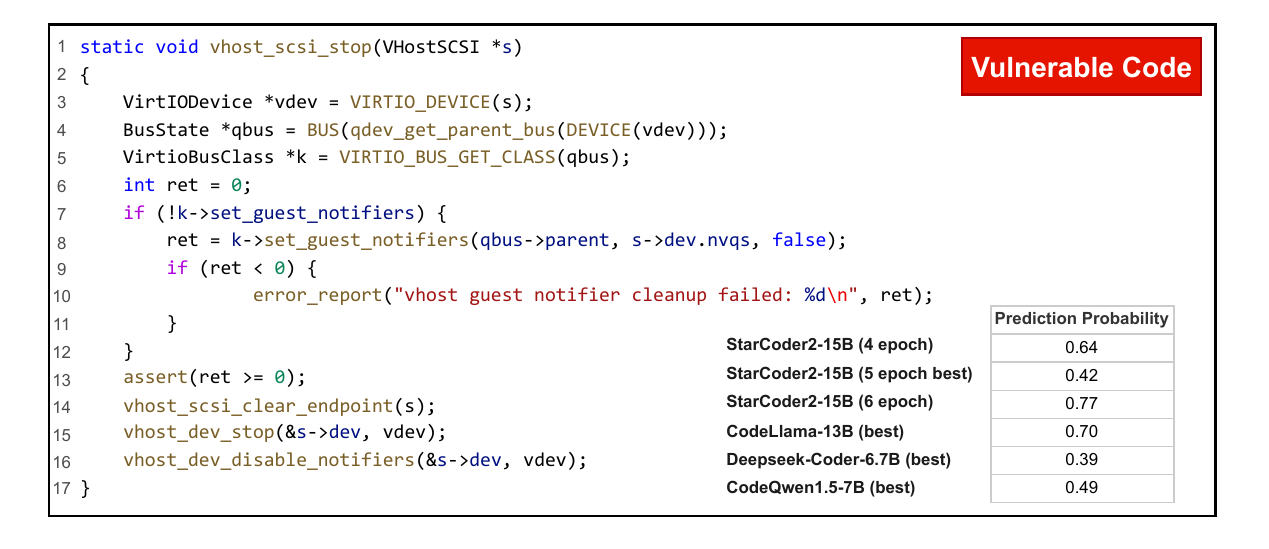}
    \caption{Examples of prediction probabilities for the same vulnerable code snippet in the Devign test set by different LLMs (different training time and model architectures). All LLMs were fine-tuned on the Devign training set for up to 10 epochs, where 'best' refers to the checkpoint selected based on the best performance on the validation set.}
    \Description{example.}
    \label{fig: example}
    % \vspace{-4mm}
\end{figure*}

% Recently, Large Language Model (LLM)-based approaches have been proposed for code vulnerability detection~\cite{shestov2024finetuning, zhou2024comparison, yang2024security, steenhoek2024comprehensive, zhou2024large, li2024llm, mao2024multi}. 
% Although these LLMs exhibit broadly consistent performance, they yield divergent detection outcomes on certain code samples. To further analyze this, we conduct a preliminary experiment. 
Recently, Large Language Model (LLM)-based approaches have been proposed for code vulnerability detection~\cite{shestov2024finetuning, zhou2024comparison, yang2024security, steenhoek2024comprehensive, zhou2024large, li2024llm, mao2024multi}. While these models demonstrate generally consistent performance, they often produce divergent detection results on specific code samples. To gain deeper insight into this phenomenon, we conducted a preliminary experiment.
% Specifically, we fine-tune multiple LLMs on the Devign dataset, including StarCoder2-15B~\cite{starcoder2}, CodeLlama-13B~\cite{codellama}, DeepSeek-Coder-6.7B~\cite{deepseek}, and CodeQwen1.5-7B~\cite{CodeQwen}. 
% We find that, although these LLMs exhibit generally consistent performance on the test set (with similar F1-score averaging around 60.8\%), they provide different prediction results for the same samples. Figure~\ref{fig: example} illustrates the varying predictions from different LLMs for the same vulnerable code snippet in the Devign~\cite{Devign} test set, where a prediction probability greater than 0.5 indicates detection of a vulnerability. 
% This suggests that while differences in training time, model architectures, and other factors do not drastically impact the overall performance of LLMs, they do affect the detection results for individual samples, implying potential complementary strengths among models in capturing diverse vulnerability patterns.
Specifically, we fine-tune several LLMs on the Devign dataset, including StarCoder2-15B~\cite{starcoder2}, CodeLlama-13B~\cite{codellama}, DeepSeek-Coder-6.7B~\cite{deepseek}, and CodeQwen1.5-7B~\cite{CodeQwen}.
While these models achieve generally consistent performance on the test set (with similar F1-score averaging around 60.8\%), their predictions often differ on the same samples.
Figure~\ref{fig: example} shows an example where the models produce varying outputs for the same vulnerable code snippet, with prediction probability greater than 0.5 indicating a detected vulnerability.
This suggests that despite comparable performance, differences in training and architecture lead to complementary strengths in capturing diverse vulnerability patterns.

\begin{figure*}
    \centering
    \begin{subfigure}[b]{0.49\linewidth}
        \centering
        \includegraphics[width=\linewidth]{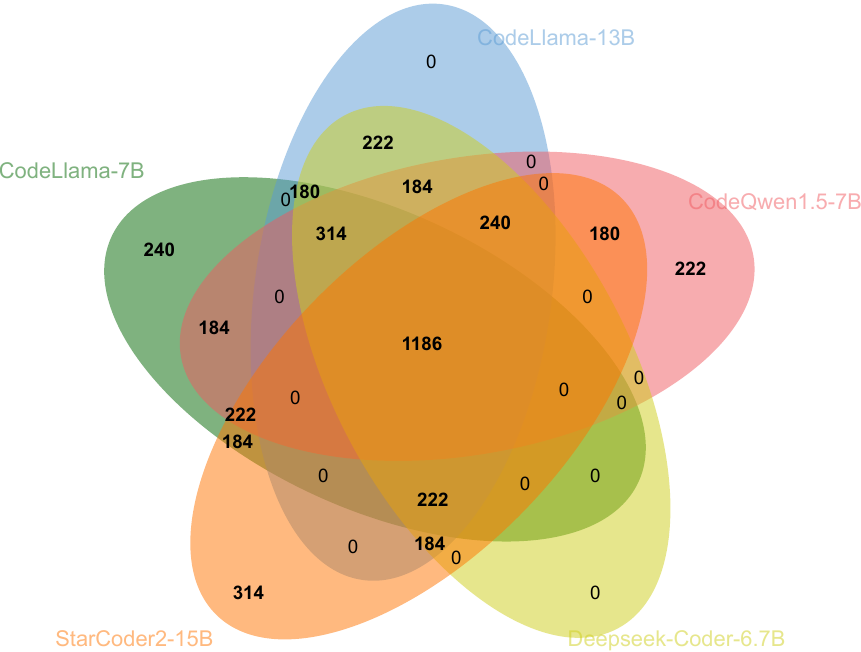}
        \caption{Different LLMs.}
        \label{fig:venn_1}
    \end{subfigure}
    \hfill
    \begin{subfigure}[b]{0.49\linewidth}
        \centering
        \includegraphics[width=\linewidth]{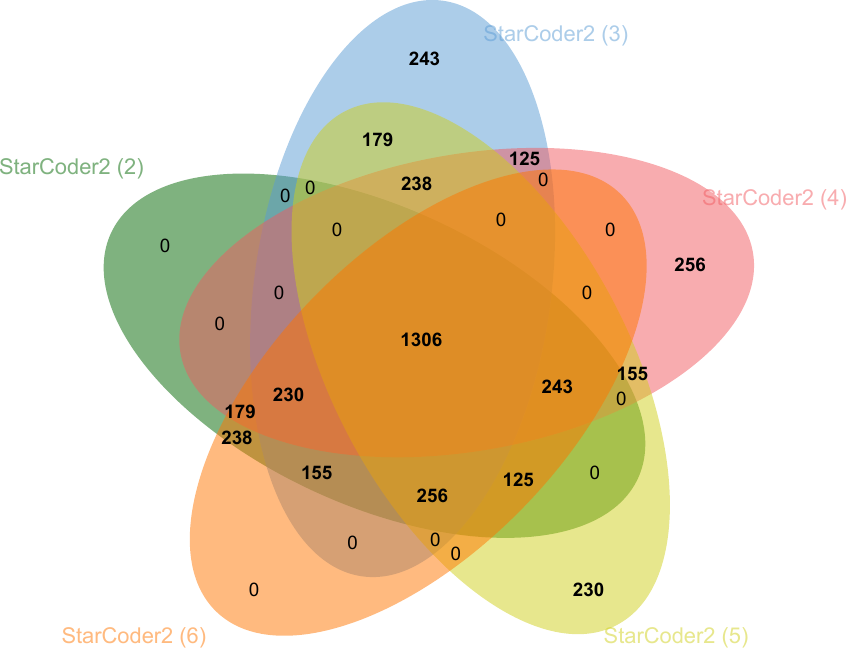}
        \caption{Different epochs for same LLM.}
        \label{fig:venn_2}
    \end{subfigure}
    % \vspace{-1em}
    \caption{Two Venn diagrams illustrating the overlap of correctly predicted samples among multiple LLMs.}
    \Description{Two Venn diagrams illustrating the overlap of correctly predicted samples among multiple LLMs.}
    \label{fig:venn}
    \vspace{-1em}
\end{figure*}

% Furthermore, to quantify this complementarity, we visualize the Venn diagram of correctly predicted samples across five LLMs (including different LLMs and different training epochs of the same LLM) in Figure~\ref{fig:venn}. Analysis of the Venn diagrams reveals that each LLM (and even different training epochs of the same LLM) identifies a distinct set of vulnerabilities. 
% For example, the heterogeneous Venn diagram (for different LLMs) in Figure~\ref{fig:venn_1} shows that, in many cases, a vulnerable sample is correctly flagged by one or more models but missed by the others. Likewise, the homogeneous Venn diagram (for multiple epochs of StarCoder2‑15B) in Figure~\ref{fig:venn_2} exhibits a similar effect, indicating that the same LLM at different training epochs also has potential complementary strengths. This phenomenon leads us to consider the following question: \textit{Can we ensemble various LLMs to achieve better performance in vulnerability detection?}
Furthermore, to quantify this complementarity, we present Venn diagrams in Figure~\ref{fig:venn} that illustrate the overlap of correctly predicted samples across five LLMs—including both different models and different training epochs of the same model. Analysis of these diagrams reveals that each LLM, and even different epochs of the same LLM, captures a unique subset of vulnerabilities.
For instance, the Venn diagram in Figure~\ref{fig:venn_1} (comparing different LLMs) demonstrates that vulnerable samples correctly identified by one or more models are often missed by others. Similarly, the Venn diagram in Figure~\ref{fig:venn_2} (comparing multiple training epochs of StarCoder2‑15B) reveals a comparable pattern, suggesting that individual training epochs also possess complementary detection capabilities.
This observation motivates the following question: \textit{Can we ensemble multiple LLMs to improve performance in vulnerability detection?}

% In the field of machine learning, ensemble learning techniques are designed to leverage the strengths of multiple models while reducing their individual weaknesses, thereby creating a more robust and accurate system~\cite{dong2020survey, sagi2018ensemble, yang2023survey}.
% The core idea of ensemble learning lies in constructing and integrating diverse models to better capture the varied characteristics of the data, thereby improving the robustness and performance of the overall system. Based on this, it is natural to apply ensemble learning techniques to improve the reliability and accuracy of predictions in LLM-based vulnerability detection by aggregating the outputs of multiple models. 
In the field of machine learning, ensemble learning techniques aim to harness the complementary strengths of multiple models while mitigating their individual weaknesses, resulting in a more robust and accurate system~\cite{dong2020survey, sagi2018ensemble, yang2023survey}. The fundamental principle of ensemble learning is to construct and combine diverse models in order to better capture the underlying complexities of the data, thereby enhancing both the robustness and overall performance of the system. Given this foundation, it is natural to apply ensemble methods to LLM-based vulnerability detection, where aggregating the outputs of multiple models can significantly improve the reliability and accuracy of predictions.

% In this paper, we first perform a comprehensive experimental evaluation to examine the effectiveness of ensemble learning in improving the performance of LLMs for vulnerability detection. Our objective is to enhance the vulnerability detection capability and robustness of LLMs by strategically leveraging their complementary strengths through ensemble learning. 
% In practice, we conduct extensive experiments on five LLMs—DeepSeek-Coder-6.7B~\cite{deepseek}, CodeLlama-7B~\cite{codellama}, CodeLlama-13B~\cite{codellama}, CodeQwen1.5-7B~\cite{CodeQwen}, and StarCoder2-15B~\cite{starcoder2}, —as well as three commonly used ensemble learning methods (i.e., Bagging~\cite{bagging}, Boosting~\cite{boosting}, and Stacking~\cite{stacking}), across three widely utilized vulnerability detection datasets ((i.e., Devign~\cite{Devign}, ReVeal~\cite{ReVeal} and BigVul~\cite{BigVul}). Drawing on recent advances in Mixture of Experts (MoE) techniques \cite{masoudnia2014mixture, cai2024survey}, we also propose a novel variant of the stacking algorithm, referred to as Dynamic Gated Stacking (DGS), tailored for vulnerability detection. To measure the performance of ensembling LLMs for vulnerability detection, we utilize four evaluation metrics, namely Accuracy, Precision, Recall, and F1-Score.
In this paper, we first perform a comprehensive empirical evaluation to investigate the effectiveness of ensemble learning in enhancing the performance of LLMs for vulnerability detection. 
% Our goal is to improve both the detection capability and robustness of LLMs by strategically leveraging their complementary strengths through ensemble techniques. 
Specifically, we perform extensive experiments using five LLMs—DeepSeek-Coder-6.7B~\cite{deepseek}, CodeLlama-7B~\cite{codellama}, CodeLlama-13B~\cite{codellama}, CodeQwen1.5-7B~\cite{CodeQwen}, and StarCoder2-15B~\cite{starcoder2}—in conjunction with three widely adopted ensemble learning methods ( i.e., Bagging~\cite{bagging}, Boosting~\cite{boosting}, and Stacking~\cite{stacking}). These experiments are conducted across three benchmark datasets (i.e., Devign~\cite{Devign}, ReVeal~\cite{ReVeal}, and BigVul~\cite{BigVul}) commonly used for vulnerability detection. Furthermore, inspired by recent developments in Mixture of Experts (MoE) techniques~\cite{masoudnia2014mixture, cai2024survey}, we introduce a variant of the stacking algorithm, termed Dynamic Gated Stacking (DGS), specifically designed for the vulnerability detection task. 
To assess the performance of LLM ensembling, we employ four standard evaluation metrics, i.e., Accuracy, Precision, Recall, and F1-Score.

In particular, we aim to explore the following Research Questions (RQs).

% \hy{any RQ that measures the robustness/consistency of the results? Like: Does ensemble learning improve the robustness of LLM-based vulnerability detection? }\zh{Section 6.2 discusses whether ensemble learning improves the inconsistency of LLMs.}
\begin{longfbox}[margin-top=0pt,margin-bottom=0pt,border-width=0pt,border-left-width=4pt,border-left-color=grey,]
\textit{\textbf{RQ1:} Does ensemble learning improve the effectiveness of LLMs in vulnerability detection?}
\end{longfbox}

To answer this RQ, we apply three traditional ensemble learning methods across five different LLMs on three datasets and compare the results against LLMs without ensemble learning. 
% \textbf{Results:} Overall, 
Experimental results show that ensemble learning enhances the performance of LLMs in vulnerability detection, with Boosting generally yielding the best average results. Surprisingly, on imbalanced datasets, all ensemble methods except Boosting fail to show improvement. Boosting demonstrates a clear advantage in Recall, particularly on imbalanced datasets. On balanced datasets, Stacking and Boosting perform similarly, while in multi-class tasks, soft-voting Bagging and Boosting show comparable performance. For Bagging, soft voting generally outperforms hard voting.

\begin{longfbox}[margin-top=0pt,margin-bottom=0pt,border-width=0pt,border-left-width=4pt,border-left-color=grey,]
\textit{\textbf{RQ2:} How do DGS perform on vulnerability detection compared to traditional Stacking algorithms?}
\end{longfbox}

In this RQ, we compare DGS with traditional stacking, aiming to explore the specific advantages of DGS over stacking in various scenarios. Experimental results show that, on the balanced dataset, stacking combined with $k$-Nearest Neighbors ($k$NN) excels in Recall and F1, while Logistic Regression and Random Forest lead in Accuracy and Precision, respectively. DGS demonstrates a clear advantage over Stacking on both the imbalanced ReVeal dataset and the multi-class BigVul task due to DGS’s integration of code features.

\begin{longfbox}[margin-top=0pt,margin-bottom=0pt,border-width=0pt,border-left-width=4pt,border-left-color=grey,]
\textit{\textbf{RQ3:} How does ensemble learning in LLMs exhibit preferences for detecting different types of vulnerabilities?}
\end{longfbox}

In this RQ, we aim to understand how different ensemble learning methods affect vulnerability detection across various vulnerability types. To this end, we extract the top ten most frequent types of vulnerabilities from the BigVul dataset and systematically evaluate the effectiveness of these ensemble methods in detecting each of these distinct types. Experimental results show that ensemble learning is effective for detecting various types of vulnerabilities, with at least one ensemble method outperforming the baseline LLMs in each case. Most ensemble methods demonstrate strong performance on memory-related vulnerabilities, while DGS shows a distinct advantage in handling multi-layered vulnerabilities.

% In summary, our major research contributions are as follows:
In summary, the primary contributions of this paper are as follows:

\setlist[itemize]{left=0pt}
\begin{itemize}
\item To the best of our knowledge, our work is the first systematic study to explore the synergy between ensemble learning and LLMs in vulnerability detection. This study not only fills a gap in the field but also offers 
practical guidance for future study.

\item Inspired by the recent MoE techniques, we propose DGS, a novel variant of the Stacking algorithm tailored for vulnerability detection.

\item We systematically investigate the impact of various factors on the performance of ensembled LLMs in vulnerability detection through extensive experiments. 

\item To support further research and advancements in this field, we have made our data and code openly accessible at \url{https://github.com/sssszh/ELVul4LLM}
\end{itemize}

We believe that our study demonstrates the potential of ensemble learning for LLM-based vulnerability detection and offers valuable insights for improving the robustness and effectiveness of LLMs.
\section{Preliminary}
\subsection{Code Vulnerability Detection}

Code Vulnerability Detection aims to identify potential security flaws in software code by analyzing code snippets (denoted as \(x \in \mathcal{X}\)) and determining whether they contain vulnerabilities. Formally, given a code snippet \(x\), a vulnerability detection tool \(f: \mathcal{X} \rightarrow \{0,1\}\) outputs a binary label \(y\), where \(y=1\) indicates a vulnerability and \(y=0\) otherwise. Such tools can be categorized into three paradigms: \textbf{static analysis} (examining code without execution)~\cite{brito2023study, kaur2020comparative, afrose2022evaluation}, \textbf{dynamic analysis} (evaluating runtime behavior)~\cite{zaazaa2020dynamic, shuai2013software}, and \textbf{deep learning (DL)-based models} (learning vulnerability patterns from labeled data)~\cite{scandariato2014predicting, li2018vuldeepecker, hin2022linevd}. Among these, DL-based models have gained prominence due to their ability to automatically learn intricate vulnerability patterns (e.g., data flow dependencies, control flow anomalies) from data, surpassing the heuristic-based limitations of traditional static and dynamic analysis tools.

DL-based approaches frame vulnerability detection as a supervised classification task. A model parameterized by \(\theta\) learns to map \(x\) to \(\hat{y} \in [0,1]\), representing the probability of \(x\) being vulnerable, by optimizing the cross-entropy loss over a training set \(D_{\text{train}}\). The objective is to find optimal parameters \(\theta^*\) that maximize the likelihood:  
\begin{equation}
\theta^* = \underset{\theta}{\arg\max} \prod_{(x,y) \in D_{\text{train}}} p(y|x,\theta).
\label{eq:likelihood}
\end{equation} 

Despite their success, conventional DL approaches often require extensive labeled datasets and may struggle with generalizing to unseen code structures. Recent advancements in LLMs have expanded their applications beyond generative tasks. Pre-trained on vast code corpora (e.g., GitHub repositories), LLMs exhibit a profound understanding of code semantics, syntax, and logical patterns, enabling them to detect subtle vulnerabilities (e.g., race conditions or memory leaks)~\cite{nam2024using}. However, LLMs inherently model \(p(x_{\text{next}}|x_{\text{prefix}})\) for sequence generation, making them suboptimal for classification tasks like vulnerability detection. Prior attempts leverage prompt-based methods (e.g., querying LLMs with instructions like ``Is this code vulnerable?'') for zero-shot detection. Yet, such methods suffer from misalignment between generative objectives and discriminative tasks, leading to unstable predictions and low performance (e.g., high false positives)~\cite{wen2024scale}. To address this, our work opts to fine-tune LLMs, adapting their generative architectures to align with the classification requirements of vulnerability detection.  

To transform an LLM into a classifier, we append a classification head (e.g., a feed-forward layer with softmax activation) atop its hidden representations. Let \(h_{\text{LLM}}(x) \in \mathbb{R}^d\) denote the contextual embedding of \(x\) produced by the LLM. The classification head projects \(h_{\text{LLM}}(x)\) into a scalar logit \(z\), followed by a sigmoid function to compute \(\hat{y}\):  
\begin{equation}
\hat{y} = \sigma\left(W \cdot h_{\text{LLM}}(x) + b\right),
\label{eq:sigmoid}
\end{equation}  
where \(W \in \mathbb{R}^{1 \times d}\) and \(b \in \mathbb{R}\) are learnable parameters. During fine-tuning, the model optimizes \(\theta_{\text{LLM}}\) and the classification head jointly by minimizing the cross-entropy loss:  
\begin{equation}
\mathcal{L} = -\sum_{(x,y) \in D_{\text{train}}} y \log \hat{y} + (1-y) \log (1-\hat{y}),
\label{eq:loss}
\end{equation}  
thereby aligning the LLM’s knowledge with the discriminative requirements of vulnerability detection.  

% \section{Large Language Models Ensembling}
% \label{sec: EL}

\subsection{Ensemble Learning}

\begin{figure*}
    \centering
    \includegraphics[width=\linewidth]{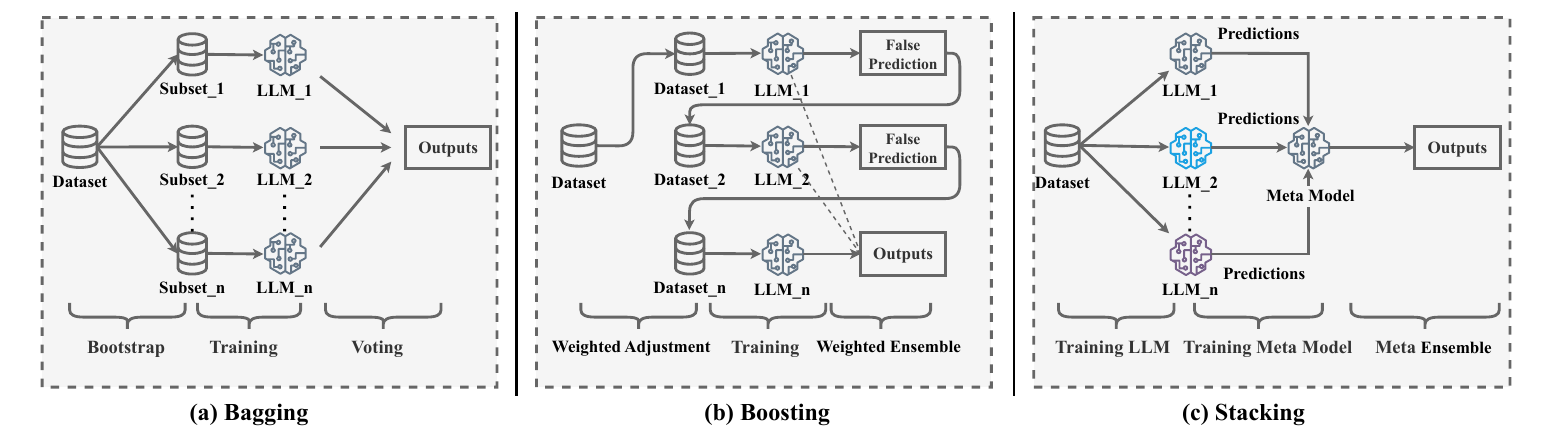}
    \caption{Overview of three traditional ensemble learning approaches we adopt: (a) Bagging, (b) Boosting, and (c) Stacking.}
    \Description{Overview.}
    \label{fig: EL}
    \vspace{-4mm}
\end{figure*}

Ensemble learning enhances the performance and robustness of machine learning models by combining predictions from multiple models, capturing diverse data features to improve generalization and predictive accuracy. Formally, given \( M \) base models \( \{f_1, f_2, \dots, f_M\} \), the ensemble prediction \( \hat{y} \) for input \( x \) is derived through an aggregation function \( \mathcal{A} \):
\begin{equation}
\hat{y} = \mathcal{A}\left(f_1(x), f_2(x), \dots, f_M(x)\right),
\end{equation}
where \( \mathcal{A} \) may represent different ensemble learning methods. As shown in Figure~\ref{fig: EL}, three classical paradigms are widely adopted: Bagging, Boosting, and Stacking.

\noindentparagraph{\textbf{\textup{Bagging.}}}

Bagging~\cite{bagging} is a seminal ensemble method that enhances model stability through parallel training of homogeneous learners. Its core innovation lies in two key mechanisms: \textit{bootstrap sampling} and \textit{aggregated prediction}.

The bootstrap process constructs $M$ subsets $\{D_1, D_2, \dots, D_M\}$ by randomly selecting $|D|$ instances \textit{with replacement} from the original dataset $D$. Statistically, each subset contains approximately 63.2\% unique samples while maintaining the original data size, thereby intentionally introducing diversity among base learners. For classification tasks, the aggregation phase employs either:
\begin{itemize}
    \item \textit{Hard voting}: Maximizes class consensus
    \begin{equation}
        \hat{y} = \arg\max_{c \in \{0,1\}} \sum_{m=1}^M \mathbb{I}\left(f_m(x) = c\right),
    \end{equation}
    
    \item \textit{Soft voting}: Averages probabilistic confidence
    \begin{equation}
        \hat{y} = \begin{cases} 
        1 & \text{if } \frac{1}{M}\sum_{m=1}^M p_m(y=1|x) \geq 0.5, \\
        0 & \text{otherwise}.
        \end{cases}
    \end{equation}
\end{itemize}

To integrate the Bagging algorithm with LLMs, we generate five bootstrapped subsets ($M=5$) through stratified sampling to preserve original class distributions. Identical large language models (LLMs) are trained in parallel, with predictions aggregated via \textbf{Bagging\_H} (hard voting) and \textbf{Bagging\_S} (soft voting). The parameter $M=5$ was selected to balance computational efficiency and variance reduction, as determined through preliminary experimentation.

\noindentparagraph{\textbf{\textup{Boosting.}}}
Boosting~\cite{boosting} constitutes a seminal paradigm in ensemble learning, distinguished by its sequential training of weak models to correct predecessors' errors. Unlike bagging's parallel structure, boosting employs an adaptive \textit{error-driven} mechanism that theoretically converts weak models into strong predictors. Among various implementations, \textbf{AdaBoost} (Adaptive Boosting)~\cite{freund1997decision} stands as the most influential and widely analyzed variant. The AdaBoost algorithm formalizes the boosting philosophy through four key phases, executed over $M$ iterations:

\begin{enumerate}
    \item \textbf{Weight initialization}: The algorithm commences by establishing an equitable starting point for all training instances. Each sample $x_i$ in the dataset $D$ receives equal influence through uniform weight distribution:
    \begin{equation}
        w_i^{(1)} = \frac{1}{N},\quad \forall i \in \{1,\dots,N\}
    \end{equation}
    
    \item \textbf{Weak model training}: At each iteration $t$, the algorithm focuses on learning the current data distribution through weighted error minimization. The weak model $f_t$ is trained to optimize the weighted misclassification error:
    \begin{equation}
        \epsilon_t = \sum_{i=1}^N w_i^{(t)} \mathbb{I}(f_t(x_i) \neq y_i)
    \end{equation}
    
    \item \textbf{Learner weight calculation}: The algorithm then computes the confidence weight $\alpha_t$ for the current learner using the closed-form solution derived from exponential loss minimization:
    \begin{equation}
        \alpha_t = \frac{1}{2}\ln\left(\frac{1-\epsilon_t}{\epsilon_t}\right) \quad 
    \end{equation}
    
    \item \textbf{Weight update}: The core adaptive mechanism manifests in dynamic weight recalibration. Misclassified samples receive exponentially increased weights according to:
    \begin{equation}
        w_i^{(t+1)} = \frac{w_i^{(t)} \cdot \exp\left(-\alpha_t y_i f_t(x_i)\right)}{Z_t}
    \end{equation}
    where $Z_t$ normalizes weights to maintain distribution.
\end{enumerate}

The final ensemble prediction combines weak models through weighted voting:
\begin{equation}
    F(x) = \text{sign}\left(\sum_{t=1}^M \alpha_t f_t(x)\right) \quad 
\end{equation}
To integrate the Boosting algorithm with LLMs, we employ it to perform multiple rounds of training on the original dataset using LLMs. In each round, a base model is trained, and the sample weights are adjusted based on the error rate from the previous round, ensuring that the next round focuses more on the misclassified samples. Finally, we integrate the predictions of all base models using a weighted voting method. In weighted voting, different weights are assigned to each base model according to its performance, and the final prediction is determined by the weighted votes.

% \noindent \textbf{Stacking}. 
\noindentparagraph{\textbf{\textup{Stacking.}}}

Stacking ~\cite{stacking} revolutionizes ensemble learning through its \textit{meta-learning} paradigm. Unlike bagging/boosting that combine homogeneous models, stacking strategically aggregates heterogeneous base models $\{f_1, \dots, f_M\}$ via a trainable meta-model $g$, effectively learning optimal combination strategies from data. Let \( \phi_m(x) = f_m(x) \) denote the prediction of the \( m \)-th model. The meta-model is trained on transformed data \( D_{\text{meta}} = \{(\phi_1(x_i), \dots, \phi_M(x_i)), y_i)\}_{i=1}^N \), which is formulated as follows:
\begin{equation}
\hat{y} = g\left(\phi_1(x), \dots, \phi_M(x)\right).
\end{equation}
To integrate the Stacking algorithm with LLMs, we use different LLMs as base models. We train several distinct LLMs on the dataset and then train a meta-model on the validation set predictions of these LLMs to learn the optimal combination of the base models' predictions. The meta-model takes the outputs of the base models as inputs and produces the final prediction. In this study, we use Logistic Regression, Random Forest, Support Vector Machine (SVM), and $k$-Nearest Neighbors ($k$NN) as meta-models to comprehensively evaluate their effectiveness in the stacking framework and analyze the impact of different meta-models on the performance of the Stacking algorithm.
\section{Study Design}

\subsection{Research Questions}
In this study, we aim to answer the following research questions:

\setlist[itemize]{left=0pt}
\begin{itemize}
\item \textbf{RQ1:} \textit{Does ensemble learning improve the effectiveness of LLMs in vulnerability detection?}

\item \textbf{RQ2:} \textit{
% How does DGS improve performance on vulnerability detection tasks compared to traditional Stacking algorithms?
How do DGS perform on vulnerability detection compared to traditional Stacking algorithms?
} 

\item \textbf{RQ3:} \textit{How does ensemble learning in LLMs exhibit preferences for detecting different types of vulnerabilities?}
\end{itemize}

These RQs investigate the effectiveness of ensemble learning for LLM-based vulnerability detection from multiple perspectives, including data balance, the types of detection tasks, and the detection of different types of vulnerabilities. In RQ1, we experiment with several ensemble learning techniques, including Bagging (Bagging\_H and Bagging\_S)~\cite{bagging}, Boosting~\cite{boosting} and Stacking~\cite{stacking}, to determine their effectiveness in enhancing LLMs' performance in code vulnerability detection. These techniques are evaluated across different conditions, considering factors such as dataset balance and the specific type of vulnerability detection task. Through RQ1, we provide practitioners with insights into selecting appropriate ensemble learning methods. In RQ2, we conduct a comparative evaluation of our proposed Stacking variant, Dynamic Gated Stacking (DGS) (see Section \ref{sec: RQ2} for more details), against traditional Stacking ensemble methods, including an assessment of various meta-model choices. RQ2 highlights the advantages of DGS over traditional Stacking methods. In RQ3, we examine the effectiveness of ensemble learning in detecting specific types of vulnerabilities using LLMs, offering insights into how ensemble methods can improve detection across different vulnerability types.

\subsection{Datasets}

In our experiments, we utilize three widely used datasets to evaluate the performance of vulnerability detection methods. Table ~\ref{tab:tabel_1} presents the statistics of the three datasets.

\setlist[itemize]{left=0pt}
\begin{itemize}
\item \textbf{Devign}~\cite{Devign} is derived from four diversified and large-scale open-source C projects: the \texttt{Linux Kernel}, \texttt{Qemu}, \texttt{Wireshark}, and \texttt{FFmpeg}. This dataset includes a total of 27,318 security-related commits, encompassing both vulnerable and non-vulnerable functions. It contains around 12,460 vulnerable functions and 14,858 non-vulnerable functions, making it a balanced dataset.

\item \textbf{ReVeal}~\cite{ReVeal} is derived from two extensive and popular open-source projects: the \texttt{Linux Debian Kernel} and \texttt{Chromium}. This dataset includes a total of 22,734 annotated functions, consisting of both vulnerable and non-vulnerable functions. It contains around 9.15\% vulnerable functions, with the rest being non-vulnerable, making it an imbalanced dataset. This dataset provides a realistic distribution of vulnerabilities, making it a valuable resource for training and evaluating deep learning models for vulnerability detection.
\item \textbf{BigVul}~\cite{BigVul} is derived from a comprehensive collection of 348 C/C++ projects from GitHub. It contains 10,900 vulnerable functions and 177,736 non-vulnerable functions, making it an imbalanced dataset. 
The primary reason for selecting this dataset is its provision of CWE vulnerability types for vulnerable functions, which facilitates the evaluation of the impact of ensemble learning on multi-class vulnerability detection in LLMs. 
We refine the dataset by removing all non-vulnerable functions and vulnerable functions lacking specific vulnerability types. This process yields 8,636 vulnerable functions across 43 distinct CWE types. Furthermore, the corresponding fixed versions of these vulnerable functions are utilized as non-vulnerable functions. As the processed dataset has a one-to-one correspondence between vulnerable functions and non-vulnerable functions, it is a balanced dataset.
\end{itemize}

\subsection{Backend LLMs}
Our experiments are based on five representative LLMs, that are open-sourced at Hugging Face\footnote{\url{https://huggingface.co/}}. 

\setlist[itemize]{left=0pt}
\begin{itemize}
\item \textbf{CodeLlama[7B\&13B]:} CodeLlama~\cite{codellama} is an LLM for code based on Llama 2, providing state-of-the-art performance among open models.

\item \textbf{DeepSeek-Coder[6.7B]:} DeepSeek-Coder~\cite{deepseek} is a multilingual LLM for source code with parameter sizes ranging from 1.3 billion to 33 billion. DeepSeek-Coder is pre-trained with a large corpus of 2 trillion tokens from 87 programming languages, including Python, C++, Java, and JavaScript. 

\item \textbf{CodeQwen1.5[7B]:} CodeQwen~\cite{CodeQwen} is pre-trained on a diverse corpus of source code from multiple programming languages, including Python, C++, and Java. This model supports extended sequence lengths of up to 8,192 tokens, facilitating the handling of complex coding tasks.

\item \textbf{StarCoder2[15B]:} StarCoder2~\cite{starcoder2} is an advanced LLM designed for code, developed by the BigCode project. It is built on the extensive Software Heritage source code archive~\cite{abramatic2018building}, covering over 600 programming languages.
\end{itemize}

\begin{table*}[!t]
\caption{Overview of the dataset statistics.
% \yao{two 8636 in the last row? check that.}\zh{Yes, the 8636 vulnerability functions and their corresponding fixed versions are non-vulnerable functions}
}
% \setlength{\tabcolsep}{1pt} % Default value: 6pt
% \footnotesize
\vspace{-1em}
% \resizebox{0.95\textwidth}{!} {
\begin{tabular}{ll|c|c|c|c}
\hline
\multicolumn{2}{l|}{\textbf{Dataset}}                 & \textbf{Task type}                   & \textbf{Non-Vulnerabilities} & \textbf{Vulnerablities} & \textbf{Ratio} \\ \hline
\multicolumn{2}{l|}{\textbf{Devign}}                  & Binary classification                & 14,858                       & 12,460                  & 1.19:1         \\ \hline
\multicolumn{2}{l|}{\textbf{ReVeal}}                  & Binary classification                & 20,494                       & 2,240                   & 9.15:1         \\ \hline
\multirow{2}{*}{\textbf{BigVul}} & \textbf{Original}  & \multirow{2}{*}{Multi-classification} & 253,096                      & 11,823                  & 21.41:1        \\ \cline{2-2} \cline{4-6} 
                                 & \textbf{Processed} &                                      & 8,636                         & 8,636 (43 CWEs)          & 1:1            \\ \hline
\end{tabular}
% }
\label{tab:tabel_1}
\vspace{-1em}
\end{table*}

\subsection{Evaluation Metrics}
Following prior studies~\cite{Devign, zhang2023vulnerability,empirical}, we select four widely used evaluation metrics for classification tasks: 

\setlist[itemize]{left=0pt}
\begin{itemize}
\item \textbf{Accuracy} examines the overall proportion of correct predictions made by a model, which is defined as follows:
\begin{equation}
\textit{Accuracy} = \frac{\textit{TP} + \textit{TN}}{\textit{TP} + \textit{TN} + \textit{FP} + \textit{FN}}
\end{equation}
where $TP$ denotes the number of correctly identified vulnerable code segments. $TN$ represents the number of correctly predicted non-vulnerable code segments. $FP$ represents the number of non-vulnerable code segments incorrectly identified as vulnerable. $FN$ represents the number of vulnerable code segments incorrectly predicted as non-vulnerable. The Accuracy reflects the model's combined capability of detecting both vulnerable and non-vulnerable code.

\item \textbf{Precision} measures the proportion of correctly identified vulnerable code among all code segments predicted as vulnerable. High precision indicates fewer false alarms in vulnerability detection. 
\begin{equation}
\textit{Precision} = \frac{\textit{TP}}{\textit{TP} + \textit{FP}}
\end{equation}

\item \textbf{Recall} quantifying the model's ability to capture true vulnerable code segments. A higher recall implies better coverage of actual vulnerabilities and fewer missed detections.
\begin{equation}
\textit{Recall} = \frac{\textit{TP}}{\textit{TP} + \textit{FN}}
\end{equation}

\item \textbf{F1-Score} represents the harmonic mean of precision and recall. It balances the trade-off between minimizing false positives (precision) and false negatives (recall), particularly critical for imbalanced vulnerability datasets.
\begin{equation}
\textit{F1-Score} = 2 \times \frac{\textit{Precision} \times \textit{Recall}}{\textit{Precision} + \textit{Recall}}
\end{equation}

\end{itemize}

For the multi-class classification task on BigVul, we use weighted evaluation metrics: 
\begin{equation}
\text{Weighted Precision} = \frac{\sum_{i=1}^{n} w_i \cdot \text{Precision}_i}{\sum_{i=1}^{n} w_i}
\end{equation}
\begin{equation}
\text{Weighted Recall} = \frac{\sum_{i=1}^{n} w_i \cdot \text{Recall}_i}{\sum_{i=1}^{n} w_i}
\end{equation} 
\begin{equation}
\text{Weighted F1} = \frac{\sum_{i=1}^{n} w_i \cdot \text{F1}_i}{\sum_{i=1}^{n} w_i}
\end{equation} 
where $w_i$ represents the weight for the $i$-th class (typically proportional to class frequency), and $n$ is the total number of classes. These weighted metrics ensure that majority classes do not dominate the evaluation, providing a more representative performance measure across all vulnerability types. In the following experiments, we will refer to Weighted Precision, Weighted Recall, and Weighted F1 as W-Precision, W-Recall, and W-F1, respectively.

\subsection{Implementation Details}

Following previous research~\cite{Devign, BigVul, sample}, we randomly split three vulnerability detection datasets into training, validation, and test sets at a ratio of 8:1:1. The training and validation sets are used for fine-tuning the LLMs. Due to the massive parameter count of LLMs, full-parameter fine-tuning demands substantial computational resources. 
Leveraging the recent parameter-efficient fine-tuning strategy for LLMs, we employ the QLoRA~\cite{dettmers2024qlora} technique, which is based on the Low-Rank Adaptation (LoRA)~\cite{hu2021lora} method. We set the rank of the low-rank matrices $r$ and $\alpha$ as 8 and 32, respectively. 
Each LLM is fine-tuned for up to 10 epochs, selecting the best-performing checkpoint on the validation set for subsequent testing on the test set. To ensure the stability and reliability of the reported results, all experiments are repeated multiple times, and the average performance is reported. All experiments are conducted on four Nvidia A100 40GB GPUs.

\section{Experimental Results and Analysis}

\subsection{RQ1: Does ensemble learning improve the effectiveness
of LLMs in vulnerability detection?}

\begin{table*}[!t]
\centering
\caption{The results of each ensemble learning approach across five different LLMs and three datasets in terms of various evaluation metrics, with w/o EL representing the absence of ensemble learning. The records that have a larger value than w/o EL are highlighted in green, with darker color indicating better performance. The best records for each metric across different models are in bold. Note that except for the stacking method, which ensembles heterogeneous LLMs, other ensemble approaches (e.g., bagging, boosting) ensemble multiple homogeneous learners derived from the same LLM (e.g., through data resampling or reweighting), rather than different LLMs.}
\setlength{\tabcolsep}{2pt} % Default value: 6pt

\resizebox{1.0\textwidth}{!} {
\begin{tabular}{llcccc|cccc|cccc}
\hline
\multirow{2}{*}{\textbf{Models}} & \multirow{2}{*}{\begin{tabular}[c]{@{}c@{}}\textbf{Ensemble learning}\\ \textbf{(EL)}\end{tabular}} & \multicolumn{4}{c|}{\textbf{Devign}} & \multicolumn{4}{c|}{\textbf{ReVeal}} & \multicolumn{4}{c}{\textbf{BigVul}} \\
 & & \textbf{Accuracy} & \textbf{Precision} & \textbf{Recall} & \textbf{F1} & \textbf{Accuracy} & \textbf{Precision} & \textbf{Recall} & \textbf{F1} & \textbf{Accuracy} & \textbf{W-Precision} & \textbf{W-Recall} & \textbf{W-F1} \\ \hline
\multirow{5}{*}{\textbf{CodeLlama-7B}} 
& \textbf{w/o EL} & 59.98 & \cellcolor{green1}56.16 & 65.73 & 60.57 & \cellcolor{green2}88.96 & \cellcolor{green2}42.77 & \cellcolor{green2}29.82 & \cellcolor{green2}35.14 & \cellcolor{green1}53.42 & 50.21 & \cellcolor{green1}53.42 & 48.15 \\
& \textbf{Bagging\_H} & \cellcolor{green1}60.54 & 55.71 & \cellcolor{green1}68.76 & \cellcolor{green1}61.55 & \cellcolor{green3}89.27 & \cellcolor{green3}44.12 & 26.32 & 32.97 & \cellcolor{green3}57.07 & \cellcolor{green3}58.20 & \cellcolor{green3}57.07 & \cellcolor{green1}50.60 \\
& \textbf{Bagging\_S} & \cellcolor{green2}61.57 & \cellcolor{green3}56.91 & \cellcolor{green2}67.25 & \cellcolor{green2}61.65 & \cellcolor{green4}\textbf{89.40} & \cellcolor{green4}\textbf{45.19} & 26.75 & 33.61 & \cellcolor{green4}\textbf{57.18} & \cellcolor{green4}\textbf{58.28} & \cellcolor{green4}\textbf{57.18} & \cellcolor{green2}50.83 \\
& \textbf{Boosting} & \cellcolor{green3}61.27 & \cellcolor{green2}56.30 & \cellcolor{green4}\textbf{70.12} & \cellcolor{green3}62.46 & 86.72 & 35.98 & \cellcolor{green4}\textbf{41.67} & \cellcolor{green4}\textbf{38.62} & 52.43 & \cellcolor{green1}50.26 & 52.43 & \cellcolor{green3}49.85 \\
& \textbf{Stacking} & \cellcolor{green4}\textbf{62.52} & \cellcolor{green4}\textbf{57.55} & \cellcolor{green4}\textbf{70.12} & \cellcolor{green4}\textbf{63.22} & 88.96 & 42.07 & 26.75 & 32.71 & \cellcolor{green2}54.77 & \cellcolor{green2}51.60 & \cellcolor{green2}54.77 & 49.27 \\ \hline

\multirow{5}{*}{\textbf{DeepSeek-Coder-6.7B}} 
& \textbf{w/o EL} & \cellcolor{green1}59.46 & \cellcolor{green1}56.42 & 65.10 & \cellcolor{green1}60.45 & \cellcolor{green4}\textbf{89.93} & \cellcolor{green4}\textbf{49.57} & \cellcolor{green1}25.44 & \cellcolor{green2}33.62 & 52.95 & 48.63 & 52.95 & 48.59 \\
& \textbf{Bagging\_H} & 59.37 & 54.72 & \cellcolor{green1}66.93 & 60.22 & 88.61 & 37.80 & 21.05 & 27.04 & \cellcolor{green2}55.56 & \cellcolor{green3}54.28 & \cellcolor{green2}55.56 & \cellcolor{green1}49.99 \\
& \textbf{Bagging\_S} & \cellcolor{green2}59.92 & 55.15 & \cellcolor{green2}68.21 & \cellcolor{green2}60.99 & 88.35 & 35.43 & 19.74 & 25.35 & \cellcolor{green4}\textbf{56.55} & \cellcolor{green4}\textbf{54.48} & \cellcolor{green4}\textbf{56.55} & \cellcolor{green2}50.62 \\
& \textbf{Boosting} & \cellcolor{green3}60.98 & \cellcolor{green2}56.24 & \cellcolor{green3}67.89 & \cellcolor{green3}61.52 & 87.81 & 38.81 & \cellcolor{green3}\textbf{37.28} & \cellcolor{green4}\textbf{38.03} & \cellcolor{green3}56.32 & \cellcolor{green1}54.22 & \cellcolor{green3}56.32 & \cellcolor{green4}\textbf{52.26} \\
& \textbf{Stacking} & \cellcolor{green4}\textbf{62.52} & \cellcolor{green3}\textbf{57.55} & \cellcolor{green4}\textbf{70.12} & \cellcolor{green4}\textbf{63.22} & 88.96 & 42.07 & 26.75 & 32.71 & \cellcolor{green1}54.77 & \cellcolor{green2}51.60 & \cellcolor{green1}54.77 & 49.27 \\ \hline

\multirow{5}{*}{\textbf{CodeQwen1.5-7B}} 
& \textbf{w/o EL} & 58.27 & 53.64 & 67.49 & 59.77 & \cellcolor{green1}88.57 & \cellcolor{green1}40.36 & \cellcolor{green1}29.39 & \cellcolor{green1}34.01 & 51.80 & 49.48 & 51.80 & 47.77 \\
& \textbf{Bagging\_H} & \cellcolor{green1}60.18 & \cellcolor{green1}55.29 & \cellcolor{green1}69.56 & \cellcolor{green1}61.61 & \cellcolor{green2}88.70 & 37.82 & 19.74 & 25.94 & \cellcolor{green1}53.01 & \cellcolor{green1}50.85 & \cellcolor{green1}53.01 & \cellcolor{green1}48.56 \\
& \textbf{Bagging\_S} & \cellcolor{green2}60.32 & \cellcolor{green2}55.38 & \cellcolor{green3}70.12 & \cellcolor{green2}61.88 & 88.43 & 35.54 & 18.86 & 24.64 & \cellcolor{green2}54.17 & \cellcolor{green3}52.99 & \cellcolor{green2}54.17 & \cellcolor{green2}49.02 \\
& \textbf{Boosting} & \cellcolor{green3}61.46 & \cellcolor{green3}56.47 & \cellcolor{green4}\textbf{70.28} & \cellcolor{green3}62.62 & \cellcolor{green3}88.79 & \cellcolor{green4}\textbf{42.62} & \cellcolor{green3}\textbf{34.21} & \cellcolor{green4}\textbf{37.96} & \cellcolor{green4}\textbf{55.97} & \cellcolor{green4}\textbf{53.87} & \cellcolor{green4}\textbf{55.97} & \cellcolor{green4}\textbf{51.97} \\
& \textbf{Stacking} & \cellcolor{green4}\textbf{62.52} & \cellcolor{green4}\textbf{57.55} & \cellcolor{green4}\textbf{70.12} & \cellcolor{green4}\textbf{63.22} & \cellcolor{green4}\textbf{88.96} & \cellcolor{green3}42.07 & 26.75 & 32.71 & \cellcolor{green3}54.77 & \cellcolor{green2}51.60 & \cellcolor{green3}54.77 & 49.27 \\ \hline

\multirow{5}{*}{\textbf{CodeLlama-13B}} 
& \textbf{w/o EL} & 60.05 & \cellcolor{green1}57.05 & 65.74 & 61.09 & \cellcolor{green1}88.57 & \cellcolor{green1}40.70 & \cellcolor{green1}30.70 & \cellcolor{green1}35.00 & 53.53 & 50.22 & 53.53 & \cellcolor{green1}49.49 \\
& \textbf{Bagging\_H} & \cellcolor{green1}61.05 & 56.03 & \cellcolor{green1}70.76 & \cellcolor{green1}62.54 & \cellcolor{green2}88.65 & 40.51 & 28.07 & 33.16 & \cellcolor{green2}56.26 & \cellcolor{green2}54.65 & \cellcolor{green2}56.26 & \cellcolor{green2}51.06 \\
& \textbf{Bagging\_S} & \cellcolor{green3}63.02 & \cellcolor{green2}57.12 & \cellcolor{green2}70.92 & \cellcolor{green3}63.28 & 88.43 & 38.10 & 24.56 & 29.87 & \cellcolor{green4}\textbf{57.60} & \cellcolor{green4}\textbf{55.98} & \cellcolor{green4}\textbf{57.60} & \cellcolor{green3}52.21 \\
& \textbf{Boosting} & \cellcolor{green4}\textbf{63.43} & \cellcolor{green3}\textbf{57.45} & \cellcolor{green4}\textbf{71.24} & \cellcolor{green4}\textbf{63.60} & 88.21 & 37.42 & \cellcolor{green4}\textbf{40.24} & \cellcolor{green4}\textbf{38.99} & \cellcolor{green3}57.30 & \cellcolor{green3}55.73 & \cellcolor{green3}57.30 & \cellcolor{green4}\textbf{54.35} \\
& \textbf{Stacking} & \cellcolor{green2}62.52 & \cellcolor{green4}\textbf{57.55} & \cellcolor{green3}70.12 & \cellcolor{green2}63.22 & \cellcolor{green4}\textbf{88.96} & \cellcolor{green3}42.07 & 26.75 & 32.71 & \cellcolor{green1}54.77 & \cellcolor{green1}51.60 & \cellcolor{green1}54.77 & 49.27 \\ \hline

\multirow{5}{*}{\textbf{StarCoder2-15B}} 
& \textbf{w/o EL} & 60.55 & \cellcolor{green1}56.88 & 65.56 & 60.91 & \cellcolor{green1}88.57 & \cellcolor{green1}40.49 & \cellcolor{green1}31.11 & \cellcolor{green1}35.19 & 52.61 & 50.87 & 52.61 & \cellcolor{green1}49.45 \\
& \textbf{Bagging\_H} & \cellcolor{green1}61.82 & \cellcolor{green2}57.11 & \cellcolor{green1}67.81 & \cellcolor{green1}62.00 & 88.52 & 40.24 & 29.82 & 34.26 & \cellcolor{green2}55.74 & \cellcolor{green2}53.77 & \cellcolor{green2}55.74 & \cellcolor{green2}52.02 \\
& \textbf{Bagging\_S} & \cellcolor{green3}63.14 & \cellcolor{green3}58.30 & \cellcolor{green2}69.40 & \cellcolor{green3}63.37 & 88.48 & 39.76 & 28.95 & 33.50 & \cellcolor{green3}56.49 & \cellcolor{green3}55.70 & \cellcolor{green3}56.49 & \cellcolor{green3}52.77 \\
& \textbf{Boosting} & \cellcolor{green4}\textbf{63.20} & \cellcolor{green4}\textbf{58.61} & \cellcolor{green4}\textbf{71.05} & \cellcolor{green4}\textbf{64.23} & 87.20 & 37.84 & \cellcolor{green4}\textbf{42.98} & \cellcolor{green4}\textbf{40.25} & \cellcolor{green4}\textbf{57.76} & \cellcolor{green4}\textbf{56.44} & \cellcolor{green4}\textbf{57.76} & \cellcolor{green4}\textbf{53.67} \\
& \textbf{Stacking} & \cellcolor{green2}62.52 & \cellcolor{green4}\textbf{57.55} & \cellcolor{green3}70.12 & \cellcolor{green2}63.22 & \cellcolor{green4}\textbf{88.96} & \cellcolor{green3}42.07 & 26.75 & 32.71 & \cellcolor{green1}54.77 & \cellcolor{green1}51.60 & \cellcolor{green1}54.77 & 49.27 \\ \hline
\end{tabular}
}
\label{tab:tabel_3}
\vspace{-1em}
\end{table*}
Table~\ref{tab:tabel_3} presents the performance metrics of various ensemble learning methods across 15 experimental instances, which include combinations of 5 different LLMs and 3 different datasets. The results indicate that most ensemble learning methods significantly improve the performance of LLMs in the task of code vulnerability detection. To provide a clearer illustration, we calculate the average ranking of each ensemble learning method across these 15 experimental instances, as shown in Table~\ref{tab:tabel_2}. The data reveal that all ensemble learning methods have superior average rankings compared to non-ensemble methods. In terms of overall performance, the Boosting algorithm excels in Recall and F1 metrics, while the Stacking algorithm achieves the best results in Accuracy, Precision and Recall. Considering the mix of balanced and imbalanced datasets, the F1 metric provides a more comprehensive evaluation of overall performance. Based on this, Boosting demonstrates the strongest overall performance with an F1 average rank of 1.60, outperforming other ensemble methods. However, Stacking also exhibits remarkable effectiveness in critical classification metrics (Accuracy, Precision and Recall), highlighting its value in scenarios prioritizing these criteria.

We analyze the reasons for the Boosting algorithm's higher average ranking from the perspective of dataset balance. As shown in Table~\ref{tab:tabel_3}, it is interesting to see that, aside from the Boosting algorithm, none of the other ensemble learning methods provide improvements on the imbalanced dataset ReVeal. In five experimental instances on this imbalanced dataset, both Bagging and Stacking algorithms demonstrated lower F1 scores compared to the baseline LLMs, with most other metrics also falling short of the baseline, except for a few cases where Bagging achieved higher accuracy. However, accuracy is not an ideal metric for evaluating vulnerability detection on imbalanced datasets. The Boosting algorithm, in contrast, is effective not only on balanced datasets but also shows significant improvements on imbalanced datasets, leading to its consistently higher average rankings across metrics in Table~\ref{tab:tabel_2} compared to other ensemble learning methods. In Section~\ref{sec: discussion}, we provide a detailed discussion of this phenomenon.

\begin{tcolorbox}[size=title,breakable]
\textit{\textbf{Finding 1:} \textcolor{black}{In general, ensemble learning improves the effectiveness of LLMs in detecting source code vulnerabilities, with Boosting and Stacking algorithms demonstrating particularly significant effectiveness. Surprisingly, most of the ensemble learning methods do not show improvement on the imbalanced ReVeal dataset, while the Boosting algorithm improves performance across both balanced and imbalanced datasets.}}
\end{tcolorbox}

When analyzing the performance of various ensemble learning methods on the imbalanced dataset ReVeal, we observe that the Boosting algorithm not only outperforms other ensemble learning methods in terms of F1 score but also demonstrates a significant advantage in Recall. In 12 out of the 15 experimental instances, the Recall score with the Boosting method was higher than that of other ensemble learning methods, with its superiority in Recall being particularly pronounced on the imbalanced datasets. Specifically, across the five LLMs tested on the imbalanced ReVeal dataset, the Boosting algorithm improved the Recall score by an average of 34.38\% relative to the baseline LLMs, whereas other ensemble learning methods tended to decrease the Recall score. In Section~\ref{sec: discussion}, we provide a detailed discussion of this phenomenon.

For balanced datasets, Table~\ref{tab:tabel_2} reveals that the Boosting method does not exhibit the same absolute advantage on the Devign dataset as it does on ReVeal. Instead, the Stacking algorithm performs comparably to Boosting on Devign, with both methods achieving the best results across different metrics. 
Out of the 20 metric comparisons (four different metrics across five LLMs on Devign), Boosting achieves the top performance in 9 instances, while Stacking leads in 11 instances. 
Therefore, for code vulnerability detection on balanced datasets, we recommend considering both Stacking and Boosting as ensemble learning methods for LLMs.

\begin{tcolorbox}[size=title,breakable]
\textit{\textbf{Finding 2:} \textcolor{black}{The Boosting algorithm demonstrates a significant advantage in Recall, particularly on the imbalanced ReVeal dataset, where it consistently outperforms other ensemble methods and achieves an average relative improvement of 34.38\% in Recall over the baseline LLMs. On the balanced dataset, Stacking and Boosting algorithms perform comparably, making both methods viable choices for code vulnerability detection.}}
\end{tcolorbox}

% \begin{table*}[!t]
% \centering
% \caption{Average rank of each ensemble learning method across five different LLMs in terms of various evaluation metrics. We use the performance of the non-ensemble learning method (denoted as w/o EL) as the baseline. For each metric, we highlight the average ranks that are better than w/o EL in green, with darker shades indicating better performance.}

% \resizebox{0.8\textwidth}{!} {

% \begin{tabular}{l|cccc}
% \hline
% \textbf{Ensemble learning} & \textbf{Accuracy} & \textbf{Precision} & \textbf{Recall} & \textbf{F1}   \\ \hline
% \textbf{w/o EL}              & 4.9      & 4.4       & 5.0    & 4.8  \\
% \textbf{Bagging\_H}        & \cellcolor{green1}3.86     & \cellcolor{green1}4.33      & \cellcolor{green1}4.13   & \cellcolor{green1}4.53 \\
% \textbf{Bagging\_S}        & \cellcolor{green3}3.13     & \cellcolor{green2}3.33      & \cellcolor{green2}3.6    & \cellcolor{green2}3.8  \\
% \textbf{Boosting}          & \cellcolor{green2}3.4      & \cellcolor{green3}3.06      & \cellcolor{green5}1.6    & \cellcolor{green5}1.6  \\
% \textbf{Stacking}          & \cellcolor{green5}2.6      & \cellcolor{green5}2.7       & \cellcolor{green3}3.4    & \cellcolor{green2}3.8  \\
% \textbf{DGS}               & \cellcolor{green4}3.06     & \cellcolor{green4}2.9       & \cellcolor{green4}2.8    & \cellcolor{green4}2.4  \\ \hline
% \end{tabular}
% }
% \label{tab:tabel_2}
% \vspace{-1em}
% \end{table*}

\begin{table*}[!t]
\centering
\caption{The average rank of each ensemble learning method across five different LLMs and three datasets in terms of multiple evaluation metrics. Specifically, for each metric in Table 2, we rank the performance of different ensemble methods (including w/o EL) under each LLM–dataset combination. We then compute the average rank of each method across all LLMs, datasets, and metrics. A lower average rank indicates better overall performance. We use the performance of the non-ensemble learning method (denoted as w/o EL) as the baseline. For each metric, we highlight the average ranks that are better than w/o EL in green, with darker shades indicating stronger performance.}

% \resizebox{0.8\textwidth}{!} {

\begin{tabular}{l|cccc}
\hline
\textbf{Ensemble learning} & \textbf{Accuracy} & \textbf{Precision} & \textbf{Recall} & \textbf{F1}   \\ \hline
\textbf{w/o EL}              & 4.07      & 4.13       & 4.07    & 4.27  \\
\textbf{Bagging\_H}        & \cellcolor{green1}3.67     & \cellcolor{green1}4.07      & \cellcolor{green1}3.60   & \cellcolor{green1}3.80 \\
\textbf{Bagging\_S}        & \cellcolor{green2}2.60     & \cellcolor{green2}2.60      & \cellcolor{green2}3.07    & \cellcolor{green2}2.73  \\
\textbf{Boosting}          & \cellcolor{green3}2.20      & \cellcolor{green3}2.20      & \cellcolor{green4}1.87    & \cellcolor{green4}1.60  \\
\textbf{Stacking}          & \cellcolor{green4}1.47      & \cellcolor{green4}1.53       & \cellcolor{green4}1.87    & \cellcolor{green2}1.87  \\ \hline
\end{tabular}
% }
\label{tab:tabel_2}
\vspace{-1em}
\end{table*}

When focusing on the multi-class vulnerability detection task, Table ~\ref{tab:tabel_2} shows that, unlike on ReVeal dataset, the Boosting algorithm does not hold an absolute advantage on the BigVul dataset. Bagging\_S demonstrates comparable performance to Boosting, with four out of five experimental instances on the BigVul dataset showing that Boosting outperforms other ensemble methods in the W-F1 metric. However, in the remaining three metrics—Accuracy, W-Precision, and W-Recall—Bagging\_S achieves the best results in four out of five instances. This suggests that while Boosting is preferable if the W-F1 metric is the primary focus, Bagging\_S should be considered for optimizing other metrics such as Accuracy, W-Precision, and W-Recall.

For the Bagging algorithm, we employ two different voting methods: Bagging\_H and Bagging\_S. As observed in Table~\ref{tab:tabel_2}, Bagging\_S consistently outperforms Bagging\_H in terms of average rankings across various metrics. Table~\ref{tab:tabel_3} further illustrates that, excluding the imbalanced dataset ReVeal, Bagging\_S achieves superior F1 scores in all 10 experimental instances compared to Bagging\_H. However, on the imbalanced ReVeal dataset, Bagging\_H outperforms Bagging\_S in all cases except for CodeLlama-7B. We hypothesize that this discrepancy arises because Bagging\_S can better utilize each model's probabilistic information on balanced datasets, allowing it to more accurately capture the confidence levels for different classes. In contrast, on imbalanced datasets, the high confidence of the models in the majority class may skew the voting results towards the majority, reducing the ability to identify the minority class, which could explain why Bagging\_S underperforms Bagging\_H on the ReVeal dataset.

\begin{tcolorbox}[size=title,breakable]
\textit{\textbf{Finding 3:} \textcolor{black}{On the BigVul dataset, while Boosting excels in the W-F1 metric, Bagging\_S outperforms in Accuracy, W-Precision, and W-Recall, making it a strong alternative for optimizing these specific metrics. Bagging\_S consistently outperforms Bagging\_H in balanced datasets, particularly in F1 scores, but does not exhibit the same performance advantage on the imbalanced ReVeal dataset, likely due to challenges in handling class imbalance effectively.}}
\end{tcolorbox}

\subsection{RQ2: How do DGS perform on vulnerability detection compared to traditional Stacking algorithms?}
\label{sec: RQ2}

The limitations of traditional Stacking algorithms observed in RQ1 - particularly their suboptimal performance on imbalanced datasets and multi-class classification scenarios - necessitate an enhanced approach for vulnerability detection tasks. Conventional Stacking architectures struggle to effectively handle the inherent complexity of source code analysis, where subtle syntactic patterns and diverse vulnerability types often coexist within imbalanced data distributions. This deficiency stems from their rigid reliance on base model predictions as the sole input for meta-learning, a design that fails to account for the rich structural and semantic features embedded in the source code itself.

Inspired by recent surge in MoE~\cite{masoudnia2014mixture, zhou2022mixture, xue2024openmoe, li2025uni} techniques, we develop a novel variant of the Stacking algorithm, refer to as Dynamic Gated Stacking (DGS). DGS is directly inspired by the gating mechanism in the MoE model. Traditional Stacking algorithms typically rely solely on the predictions of base models to train the meta-model. However, for the complex task of vulnerability detection, depending only on these predictions may not be sufficient to capture the intricate nature of source code. Therefore, DGS incorporates a gating mechanism similar to that found in the MoE model. It not only takes into account the predictions of the base models but also integrates the features of the source code itself. Specifically, unlike traditional Stacking, which uses the predictions of base models to produce the final output, DGS takes the source code as input and requires the meta-model to predict which LLM is most appropriate to handle the given code. During training, DGS considers both the features of the source code and the predictions of the base models to train the meta-model. This approach enables DGS to better select the most suitable base models based on the specific characteristics of the code being analyzed, thereby enhancing the accuracy and reliability of vulnerability detection.

Table~\ref{tab:tabel_4} presents the results of DGS and Stacking ensemble learning methods using different meta-models across three different datasets. When focusing on the performance on the balanced dataset Devign, both Stacking and DGS outperform the baseline LLMs across most metrics. Notably, the Stacking with $k$NN as the meta-model achieves the best results in F1 and Recall, while the Stacking methods using Logistic Regression  and Random Forest as meta-models excel in Accuracy and Precision, respectively.

We offer the following analysis for these observations. The $k$NN model makes classification decisions based on the nearest neighbors, which gives it an advantage in identifying boundary samples, making it highly effective in recognizing minority class samples and thus improving Recall and F1. Although Devign is a balanced dataset, this can also be explained by $k$NN’s performance on the ReVeal dataset, where it achieved the highest Recall among the four different meta-models. Logistic Regression  generally performs well when the overall data distribution is uniform and linearly separable, making it effective in optimizing overall Accuracy. Random Forest, which is based on multiple decision trees, has a strong capability for class differentiation and tends to produce fewer false positives, hence excelling in improving Precision. Therefore, for balanced datasets, if the objective is to optimize Recall and F1, $k$NN is a suitable choice as the meta-model; for Accuracy, Logistic Regression  is recommended; and for Precision, Random Forest should be preferred.

\begin{tcolorbox}[size=title,breakable]
\textit{\textbf{Finding 4:} \textcolor{black}{On the balanced Devign dataset, Stacking with $k$NN as the meta-model excels in Recall and F1, while Logistic Regression  and Random Forest meta-models excel in Accuracy and Precision, respectively, indicating that different meta-models can be optimized for specific performance metrics.}}
\end{tcolorbox}

When focusing on the imbalanced dataset ReVeal and the multi-class vulnerability detection task BigVul, Table~\ref{tab:tabel_4} reveals that all four Stacking ensemble methods with different meta-models fail to improve F1 and Recall on the imbalanced ReVeal dataset, with performance actually declining significantly. Although the Stacking method using Random Forest as the meta-model achieves the best results in Accuracy and Precision, Accuracy is not a reliable metric for evaluating the model's code vulnerability detection capabilities on imbalanced datasets, whereas F1 provides a more meaningful assessment. Similarly, in the multi-class vulnerability detection task, none of the Stacking methods outperformed the baseline LLMs in terms of F1.

% \usepackage{multirow}
\begin{table*}[!t]
\centering
\vspace{-1em}
\caption{The results of various metrics across three datasets for Stacking algorithms with different meta-models and the DGS algorithm. Results that outperform all baseline LLMs are highlighted in green, with darker shades indicating better performance.}
\setlength{\tabcolsep}{2pt} % Default value: 6pt

\resizebox{1.0\textwidth}{!} {
\begin{tabular}{llcccccccccccc}
\hline
\multicolumn{2}{c}{}                    & \multicolumn{4}{c}{\textbf{Devign}}                                        & \multicolumn{4}{c}{\textbf{ReVeal}}                                        & \multicolumn{4}{c}{\textbf{BigVul}}                                        \\
\multicolumn{2}{c}{}                    & \textbf{Accuracy}            & \textbf{Precision}              & \textbf{Recall}              & \textbf{F1}             & \textbf{Accuracy}            & \textbf{Precision}              & \textbf{Recall}              & \textbf{F1}             & \textbf{Accuracy}            & \textbf{W-Precision}            & \textbf{W-Recall}            & \textbf{W-F1}           \\ \hline
\multicolumn{2}{l}{\textbf{CodeLlama-7B}}        & 59.98          & 56.16          & 65.73          & 60.57          & 88.96          & 42.77          & 29.82          & 35.14          & 53.42          & 50.21          & 53.42          & 48.15          \\
\multicolumn{2}{l}{\textbf{DeepSeek-Coder-6.7B}} & 59.46          & 56.42          & 65.10          & 60.45          & 89.93          & 49.57          & 25.44          & 33.62          & 52.95          & 48.63          & 52.95          & 48.59          \\
\multicolumn{2}{l}{\textbf{CodeQwen1.5-7B}}         & 58.27          & 53.64          & 67.49          & 59.77          & 88.57          & 40.36          & 29.39          & 34.01          & 51.08          & 49.48          & 51.80          & 47.77          \\
\multicolumn{2}{l}{\textbf{CodeLlama-13B}}       & 60.05          & 57.05          & 65.74          & 61.09          & 88.57          & 40.70           & 30.70           & 35.00          & 53.53          & 50.22          & 53.53          & 49.49          \\
\multicolumn{2}{l}{\textbf{StarCoder2-15B}}      & 60.55          & 56.88          & 65.56          & 60.91          & 88.57          & 40.49          & 31.11          & 35.19          & 52.61          & 50.87          & 52.61          & 49.45          \\ \hline
\multirow{4}{*}{\textbf{Stacking}}      & \textbf{$k$NN}    & \cellcolor{green2} 62.52 & \cellcolor{green1} 57.55 & \cellcolor{green4} \textbf{70.12} & \cellcolor{green3} \textbf{63.22} & 88.96          & 42.07          & 26.75          & 32.71          & 51.85          & 42.36          & 51.85          & 43.54          \\
                               & \textbf{Logistic Regression}     & \cellcolor{green4} \textbf{63.25}          & \cellcolor{green2} 61.78          & 52.43          & 56.72          & \cellcolor{green3} 90.37          & \cellcolor{green2} 59.57          & 12.28          & 20.36          & 49.25          & 25.37          & 49.25          & 33.48          \\
                               & \textbf{Random Forest}     & \cellcolor{green3} 62.96          & \cellcolor{green4} \textbf{62.97}          & 47.01          & 53.83          & \cellcolor{green4} \textbf{90.55} & \cellcolor{green4} \textbf{60.00} & 17.10           & 26.62          & \cellcolor{green2} 54.77          & \cellcolor{green2} 51.60           & \cellcolor{green2} 54.77          & 49.27          \\
                               & \textbf{SVM}    & \cellcolor{green1} 62.77          & \cellcolor{green3} 61.85          & 47.22          & 53.55          & \cellcolor{green2} 90.21          & \cellcolor{green2} 59.84          & 18.01          & 27.69          & 50.41          & 27.72          & 50.41          & 34.39          \\ \hline
\multicolumn{2}{c}{\textbf{DGS}}                 & \cellcolor{green5} 61.75          & 56.81          & \cellcolor{green3} 69.80          & \cellcolor{green3} 62.64          & 88.87          & 42.60          & \cellcolor{green4} \textbf{31.58} & \cellcolor{green4} \textbf{36.27} & \cellcolor{green4} \textbf{55.10} & \cellcolor{green4} \textbf{52.65} & \cellcolor{green4} \textbf{55.10} & \cellcolor{green4} \textbf{51.51} \\ \hline
\end{tabular}
}
\label{tab:tabel_4}
\vspace{-1em}
\end{table*}

In contrast, DGS achieves the best F1 and Recall results on the imbalanced ReVeal dataset and demonstrates superior performance across all metrics in the multi-class vulnerability detection task. Unlike traditional Stacking, DGS incorporates additional source code features to train the meta-model, which likely aids in better capturing the complex code structures and semantic information. This approach enhances the model's ability to recognize minority class samples, which are typically more challenging to detect on the imbalanced ReVeal dataset. By integrating code features, DGS appears to improve sensitivity to minority classes and handles the complexities of multi-class scenarios with greater accuracy.

\begin{tcolorbox}[size=title,breakable]
\textit{\textbf{Finding 5:} \textcolor{black}{DGS demonstrates a clear advantage over traditional Stacking on both the imbalanced ReVeal dataset and the multi-class BigVul task. This superiority is likely due to DGS's integration of source code features, which enhances its ability to capture complex code structures and improve recognition of minority class samples.}}
\end{tcolorbox}

\begin{figure}
    \centering
    \begin{subfigure}[b]{\linewidth}
        \centering
        \includegraphics[width=.8\linewidth]{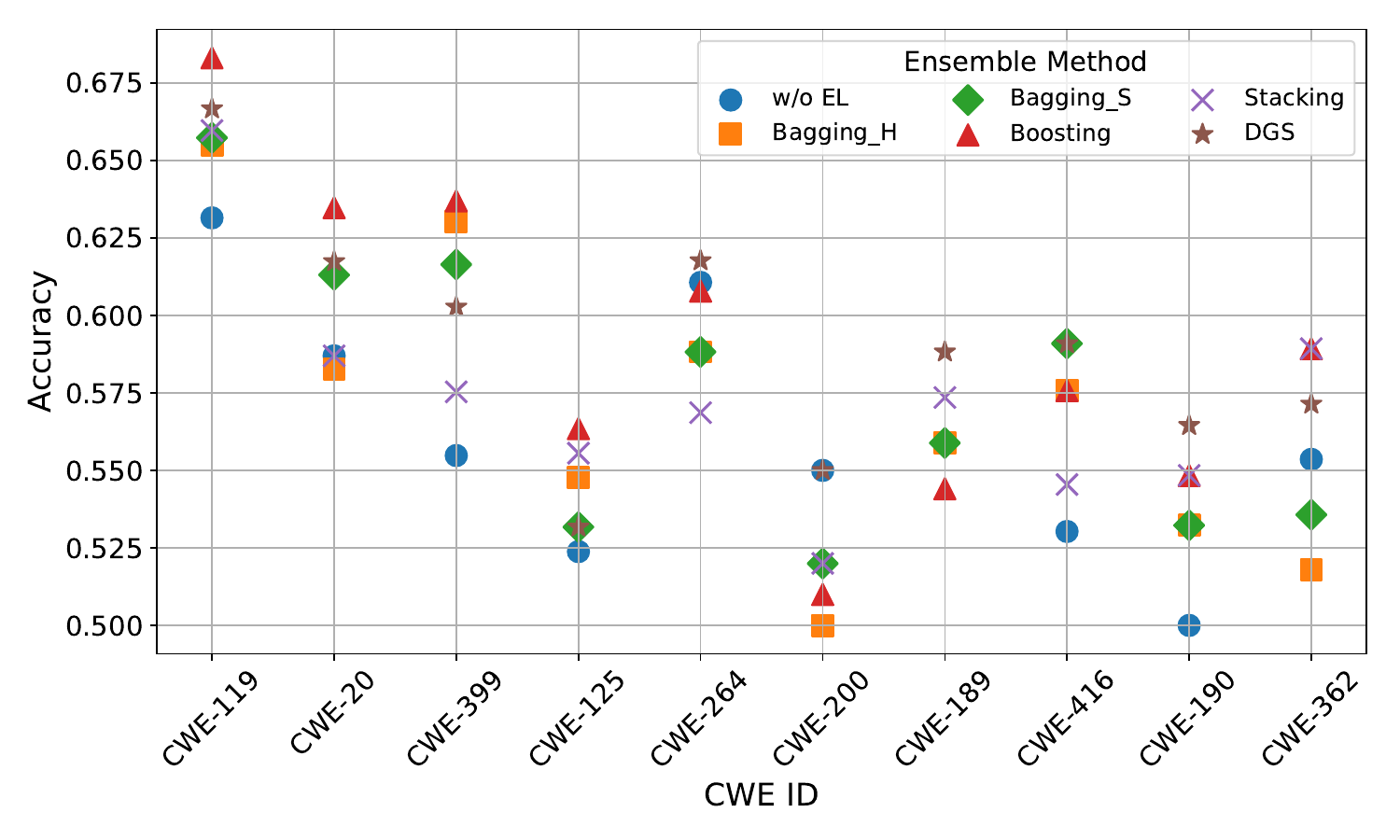}
        \caption{CodeQwen1.5-7B}
        \label{fig:case_1}
    \end{subfigure} 

    \begin{subfigure}[b]{\linewidth}
        \centering
        \includegraphics[width=.8\linewidth]{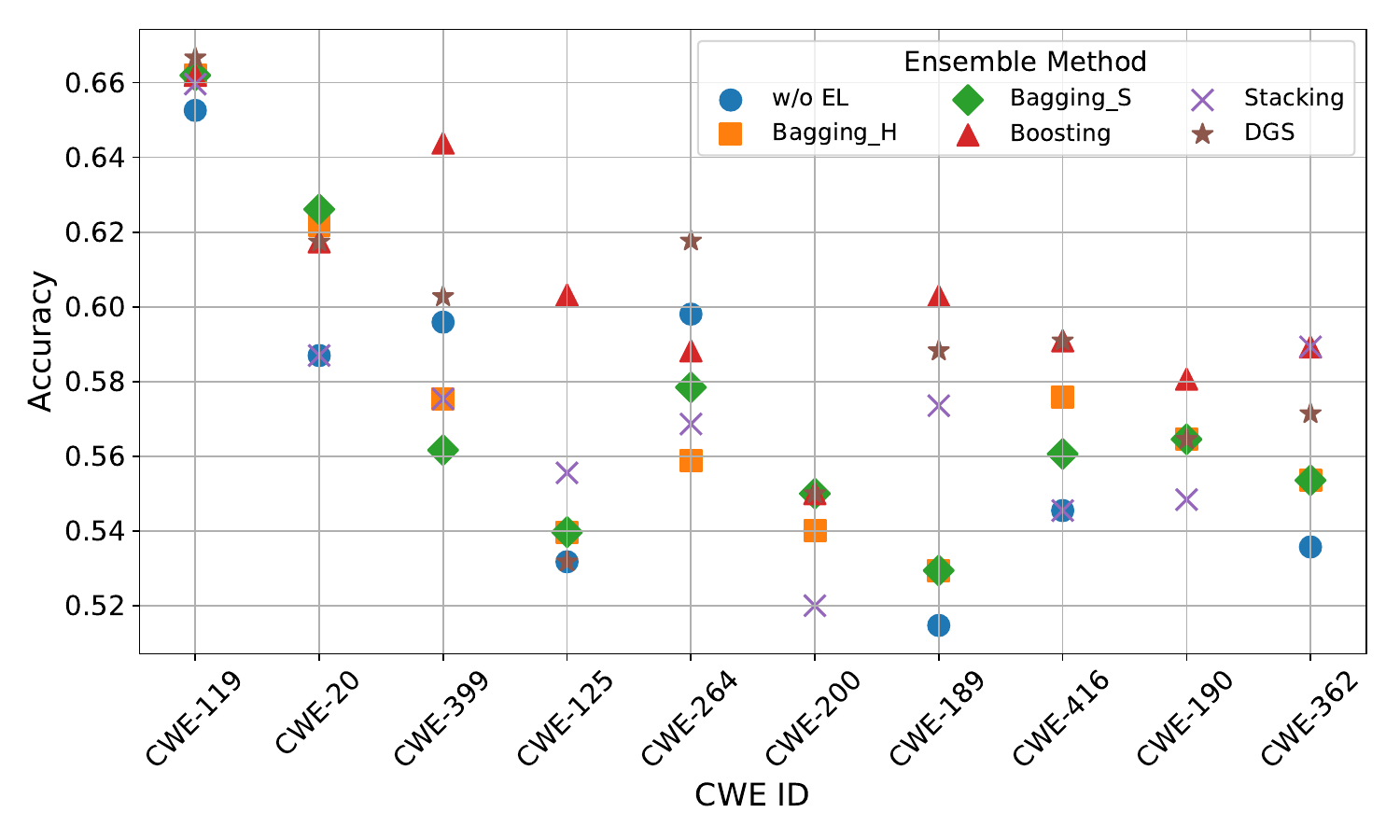}
        \caption{CodeLlama-13B}
        \label{fig:case_2}
    \end{subfigure} 

    \caption{Performance of different ensemble methods across various CWEs on CodeQwen1.5-7B and CodeLlama-13B.}
    \label{fig:cwe}
    \vspace{-1em}
\end{figure}

\subsection{RQ3: How does ensemble learning in LLMs exhibit preferences for detecting different types of vulnerabilities?}

To investigate the differences in how various ensemble learning methods detect specific types of vulnerabilities, we extract the top 10 most frequent CWEs from the BigVul dataset to create 10 subsets. We then evaluate different ensemble learning methods on these subsets. Given that the ratio of vulnerable to non-vulnerable functions in these subsets is 1:1, we use accuracy as the evaluation metric. Figure~\ref{fig:cwe} illustrates the detection performance of different ensemble learning methods across these 10 CWEs using two different models, CodeQwen1.5-7B and CodeLlama-13B.

Based on the results shown in Figure~\ref{fig:cwe}, we observe that ensemble learning methods are generally effective across different types of vulnerability detection, with at least one ensemble method outperforming the baseline LLMs for each CWE. Additionally, we identify similar patterns across the two models, where most ensemble learning methods demonstrate a consistent advantage in detecting CWE-119, CWE-120, CWE-125, and CWE-416. Upon reviewing the CWE community definitions, we find that CWE-119 involves improper restriction of operations within the bounds of a memory buffer; CWE-120 refers to a classic buffer overflow vulnerability, typically occurring when input data is copied into a buffer without proper size checks; CWE-125 relates to out-of-bounds read vulnerabilities, where a program reads data beyond its allocated buffer or memory space; and CWE-416 involves use-after-free vulnerabilities, where a program continues to use a pointer or reference after freeing the memory. These four CWEs are all related to memory operations, suggesting that ensemble learning methods, particularly the Boosting algorithm, are well-suited for detecting vulnerabilities associated with memory handling.
\vspace{-1mm}
\begin{tcolorbox}[size=title,breakable]
\textit{\textbf{Finding 6:} \textcolor{black}{Ensemble learning methods, particularly in detecting memory-related vulnerabilities such as CWE-119, CWE-120, CWE-125, and CWE-416, consistently outperform baseline LLMs across different models, with Boosting generally showing the best overall performance.}}
\end{tcolorbox}
\vspace{-1mm}

When evaluating the performance of ensemble learning methods across various vulnerability types, we found that most methods underperform compared to the baseline LLMs on CWE-264, whereas our proposed DGS method excels and surpasses the baseline. CWE-264 involves permissions and access controls, typically spanning multiple system layers like file permissions, database access, and user authentication. Detecting such multi-layered vulnerabilities requires diverse interpretations from different LLMs, which may vary in their understanding. Since Bagging and Boosting algorithms use the same base models, they lack an advantage in this scenario. Although Stacking uses different LLMs, CWE-264's complex permissions and access control logic are often closely tied to specific code structures and contexts. By incorporating source code features, DGS captures these contextual details, enabling more accurate detection of permission-related vulnerabilities.

\begin{tcolorbox}[size=title,breakable]
\textit{\textbf{Finding 7:} \textcolor{black}{For multi-layered vulnerabilities that require different interpretations from various LLMs, the traditional ensemble learning methods perform poorly. However, DGS demonstrates a clear advantage in handling these complex vulnerabilities.}}
\end{tcolorbox}
\section{Discussion}
\label{sec: discussion}

\subsection{Why are boosting algorithms effective in imbalanced datasets and particularly good at enhancing recall?}

In RQ1, we observe that most ensemble learning methods fail on the imbalanced ReVeal dataset, except for the Boosting method, which demonstrates a significant advantage in improving the Recall metric. We provide a detailed analysis of this phenomenon. We hypothesize that the ineffectiveness of other ensemble methods on ReVeal is primarily due to the severe impact of data imbalance on the performance of LLMs in vulnerability detection. Bagging and Stacking methods are not capable of mitigating the adverse effects of this imbalance. In contrast, Boosting adjusts the sample weights in each iteration based on the classification errors of the previous round. We speculate that in the imbalanced ReVeal dataset, minority class samples—specifically, the vulnerable code segments—if misclassified, have their weights increased in subsequent iterations, thereby directing the model's attention toward these hard-to-classify minority samples.

\begin{figure}[!t]
    \centering
    \includegraphics[width=.9\linewidth]{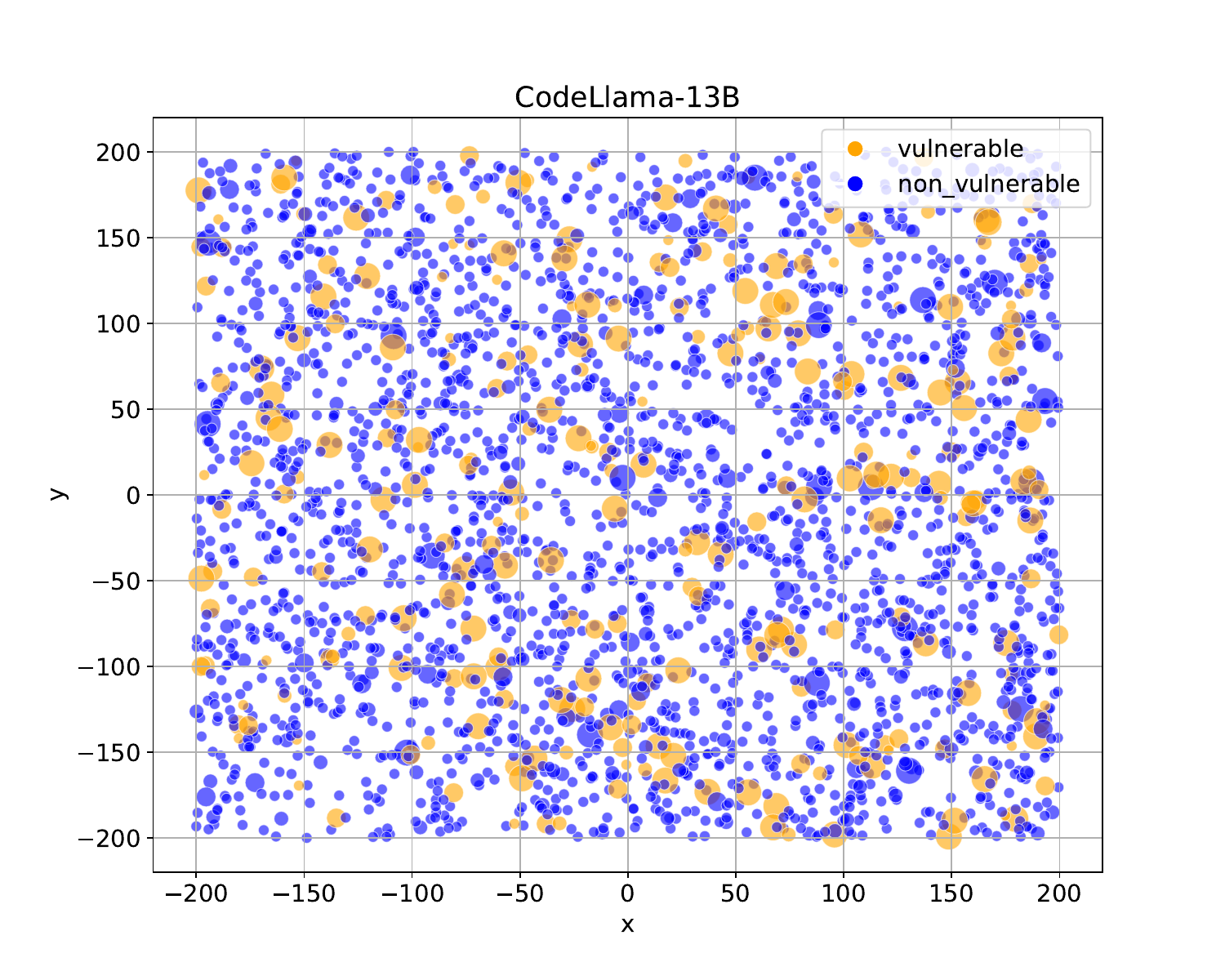}
    \vspace{-1em}
    \caption{Training sample weights during boosting process on CodeLlama-13B.}
    % \vspace{-3mm}
    \label{fig: boost}
    \vspace{-1em}
\end{figure}

We empirically validate this hypothesis by visualizing the training sample weights through their random distribution in a space. Figure~\ref{fig: boost} illustrates the training sample weights when training the CodeLlama-13B model on ReVeal using Boosting\footnote{Due to space limitations, we provide one model example. The X and Y axes in the figure represent a random spatial distribution without specific semantic meaning, serving solely to visualize the relative positions of samples.}. Different colors represent vulnerable and non-vulnerable samples, and the size of the points represents the weight of the training samples. As observed, the minority vulnerable function samples have larger weights during the training process, confirming our hypothesis. LLMs struggle to classify these minority vulnerable samples, and Boosting progressively increases the weights of misclassified samples. This means that LLMs gradually reduce their bias toward the majority class during training, balancing their ability to handle different classes. Since Boosting tends to focus more on misclassified samples, particularly vulnerable functions misclassified as non-vulnerable, it effectively reduces the number of false negatives, thereby improving Recall.

\subsection{How does ensemble learning leverage complementary strengths of LLMs in vulnerability detection?}

\begin{figure}[!t]
    \centering
    \includegraphics[width=0.8\linewidth]{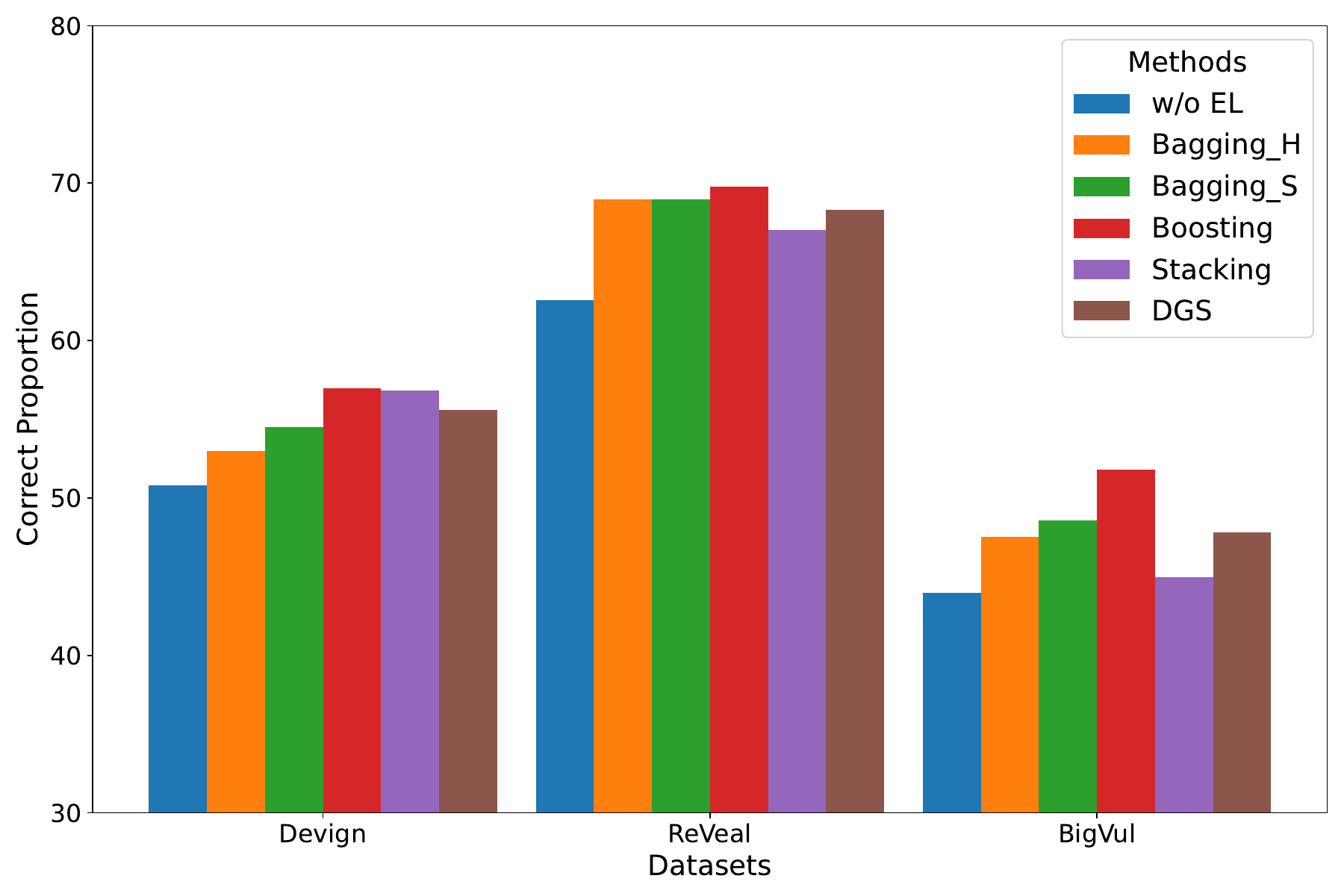}
    \caption{
    Proportions of correct predictions on divergent samples (i.e., samples with conflicting predictions across LLM instances) across methods and datasets, using CodeLlama-13B as backend. 
    }
    \vspace{-1mm}
    \label{fig: inconsistent}
    \vspace{-1mm}
\end{figure}
While evaluation metrics demonstrate the performance enhancement brought by ensemble learning, it is essential to investigate whether this improvement originates from effectively harnessing the complementary strengths of diverse LLM instances. To explore this, we analyze predictions from five distinct epochs of CodeLlama-13B—all trained under identical conditions—across three datasets.
We focus specifically on divergent samples, as these cases highlight scenarios where individual models exhibit differing capabilities and perspectives.
Our analysis identifies 1,683 divergent samples in Devign (out of 2,732 test samples), 243 in ReVeal (out of 2,274), and 807 in BigVul (out of 1,726).
We then compare the proportion of correct predictions between standalone models (baseline) and ensemble methods on these divergent samples. This metric reflects how effectively ensemble learning synthesizes diverse model perspectives to improve decision-making.

As shown in Figure~\ref{fig: inconsistent}, ensemble methods consistently outperform the baseline across all datasets, achieving higher accuracy on divergent samples. For the ReVeal dataset, while ensemble learning demonstrates improved accuracy on divergent samples, the overall classification metrics show limited enhancement. This observation stems from the inherent class imbalance in the data: LLMs exhibit a strong bias toward predicting the majority class, leading to fewer divergent samples (only 243 out of 2,274 test samples). Consequently, the influence of these samples on aggregated evaluation metrics is marginal. However, the higher correct prediction rate on divergent samples—achieved by ensemble methods—reveals their ability to reconcile conflicting predictions and leverage complementary model insights, even in imbalanced scenarios. This outcome highlights the dual role of ensemble learning: while its impact on holistic metrics may be constrained by data imbalance, it remains a powerful strategy to harness model diversity for resolving ambiguous cases and improving decision robustness.

\subsection{Implications}

\mypara{We recommend prioritizing the use of Boosting algorithms for vulnerability detection in scenarios where data class distribution is imbalanced} 
In our study, Boosting algorithms consistently outperform other ensemble methods on the ReVeal dataset, which has an imbalanced class distribution. Their ability to enhance recall and F1 scores indicates that Boosting algorithms are more effective at identifying minority class samples. Therefore, we recommend prioritizing the use of Boosting algorithms for vulnerability detection tasks in scenarios where class distribution is imbalanced.

\mypara{We recommend exploring new methods that integrate domain-specific features into ensemble techniques to enhance the accuracy of vulnerability detection} 
Given the success of DGS in leveraging source code features, future research can focus on developing new ensemble approaches that incorporate similar or more advanced domain-specific features, such as syntax, structural information of code, or vulnerability-related information. These techniques have the potential to improve model sensitivity and accuracy, particularly in challenging datasets with complex structures.

\mypara{We recommend further investigation into the specific types of vulnerabilities that ensemble learning methods are best suited to detect} 
Our findings indicate that ensemble methods, particularly Boosting, perform exceptionally well in detecting memory-related vulnerabilities such as CWE-119 and CWE-120. However, we only conduct experiments on a limited set of CWEs. Future research should focus on exploring a broader range of vulnerability types to identify where ensemble methods offer the greatest advantages. This will enable a more targeted and effective application of these techniques across different vulnerability categories.

\mypara{We recommend investigating the strengths of different LLMs for detecting specific types of vulnerabilities and exploring the integration of ensemble learning with MoE} 
In RQ3, we also observe that different LLMs may exhibit varying capabilities in detecting different types of vulnerabilities, indicating that different LLMs can act as specialized experts. Therefore, we recommend further exploration of combining ensemble learning with MoE in future research. For example, LLMs fine-tuned on datasets specific to certain vulnerability types may excel in identifying those particular vulnerabilities. By leveraging ensemble learning to harness the strengths of each LLM, this combination could offer a more targeted and effective approach to vulnerability detection.

\subsection{Computational Costs of Ensemble Learning}
While our experimental results demonstrate the effectiveness of ensemble learning in improving LLM-based vulnerability detection, it is critical to address the associated computational costs. Ensemble methods inherently require training or fine-tuning multiple LLMs and combining their predictions, which introduces significant computational overhead compared to single-model baselines.

\mypara{Training Costs} Methods like Bagging and Boosting involve training multiple LLM instances. For example, Bagging requires training five LLMs on bootstrapped datasets, while Boosting iteratively trains models with adjusted sample weights. This multiplies the computational resources (e.g., GPU hours) by the number of base models. Stacking and DGS further introduce meta-model training, adding moderate overhead depending on the complexity of the meta-learner.

\mypara{Inference Costs} During inference, ensemble methods require running multiple LLMs in parallel, increasing latency and memory usage. For instance, Bagging\_S averages predictions from five LLMs, while DGS dynamically routes inputs to selected models. While DGS reduces redundant computations compared to traditional Stacking, its gating mechanism still introduces additional processing steps.

Optimizing ensemble efficiency remains an open challenge. The choice of ensemble method should consider both performance gains and resource constraints. Boosting, while effective on imbalanced data, demands sequential training iterations, making it less scalable for large-scale deployments. In contrast, Bagging allows parallel training but requires maintaining multiple model instances. In summary, while ensemble learning enhances detection robustness, practitioners must carefully evaluate its cost-benefit trade-offs in resource-constrained environments. Our findings highlight the need for further research into efficient ensemble architectures tailored for LLM-based security applications.
\section{Threats to Validity}
% \noindent \textbf{Internal Validity.} 
\mypara{Internal Validity}
In this study, a potential threat arises from the hyperparameter tuning of fine-tuned LLMs. Despite the use of parameter-efficient tuning techniques, the process of optimizing hyperparameters for these models is resource-intensive. Consequently, the hyperparameters employed may not be optimal, potentially impacting the performance of the fine-tuned LLMs. However, the primary objective of this study is not to maximize the performance of individual LLMs. Instead, our focus is on exploring whether ensemble learning methods can improve the accuracy and stability of LLM-based vulnerability detection. To address this threat, we standardize the hyperparameters across all fine-tuned LLMs, thereby ensuring that the evaluation of ensemble learning's impact on vulnerability detection was conducted under consistent conditions.

% \noindent \textbf{External Validity.}
\mypara{External Validity}
External validity threats pertain to the generalizability of our findings. Our study focuses on a limited set of ensemble learning methods, datasets, and LLMs. Therefore, these findings may not generalize to other ensemble learning methods, datasets, and LLMs. To mitigate this threat, we select the most commonly used ensemble learning methods in machine learning, consider the balance of the datasets and the types of tasks, and choose the five most popular LLMs. However, these choices do not fully address the threat. We encourage future research to explore a broader range of ensemble learning methods, datasets, and LLMs.
\section{Related Work}
\subsection{LLM-based Code Vulnerability Detection}

LLMs, which are language models based on the Transformer~\cite{vaswani2017attention}, have profoundly transformed the field of generative artificial intelligence, demonstrating considerable capabilities in text generation~\cite{GPT-4}, speech synthesis~\cite{anastassiou2024seed}, and video creation~\cite{liu2024sora}. Specially, the LLMs have also shown significant performance in code-related tasks, such as code generation~\cite{code-gen-0,code-gen-1,code-gen-2, aiXcoder-7B}, program repair~\cite{code-repair-0,code-repair-1}, and code completion~\cite{code-comp-0,code-comp-1, aiXcoder-7B-v2}. Numerous studies have also been proposed for the task of vulnerability detection~\cite{shestov2024finetuning, zhou2024comparison, yang2024security, steenhoek2024comprehensive, zhou2024large, li2024llm, mao2024multi}. 
For instance, Shestov et al.~\cite{shestov2024finetuning} investigated the fine-tuning of LLMs on vulnerability detection datasets, demonstrating that LLMs perform well in these tasks. 
Zhou et al.~\cite{zhou2024comparison} conducted a more comprehensive analysis, examining the impact of different programming languages and LLM adaptation techniques (such as prompt-based and fine-tuning approaches) on vulnerability detection. 
Mao et al.~\cite{mao2024multi} proposed a multi-role consensus approach, where LLMs simulate various roles in the code review process to detect vulnerabilities, thereby enabling a more comprehensive identification and classification of vulnerabilities in code. 
Additionally, Li et al.~\cite{li2024llm} combined LLMs with static analysis tools for vulnerability detection, showing that LLMs often outperform many advanced static analysis tools. 
These studies collectively highlight the great potential of LLMs in the field of vulnerability detection.

\subsection{Ensemble Learning for Software Engineering}
With the advancement of machine learning technologies, ensemble learning has emerged as an effective means to enhance the robustness and stability of models, finding wide application across various software engineering tasks. By aggregating the predictions of multiple models, ensemble learning significantly improves the performance of individual models in terms of accuracy, stability, and generalization capabilities. Remarkable achievements have been made in applying ensemble learning to tasks such as software defect prediction~\cite{software-defect-0,software-defect-1, software-defect-2}, software cost estimation~\cite{software-cost}, software bug classification~\cite{software-bug}, software quality prediction~\cite{software-quality}, and software change prediction~\cite{software-change}. For instance, Dong et al.~\cite{software-defect-0} conducted an extensive study exploring the performance of various machine learning and deep learning algorithms in software defect prediction. They utilized three classical ensemble learning methods—Bagging, Boosting, and Stacking—to enhance the prediction performance. Alhazmi et al.\cite{software-cost} conducted a comprehensive study on improving software effort prediction accuracy using ensemble learning methods. Their study found that ensemble learning methods can significantly improve the accuracy and stability of models in software effort prediction, with the Bagging algorithm being particularly effective. Ceran et al.~\cite{software-quality} conducted a study aimed at improving the accuracy of software quality prediction through machine learning-based ensemble methods. The study's results indicate that ensemble methods significantly enhance the accuracy of software quality prediction compared to individual machine learning algorithms. While previous studies primarily focus on the application and evaluation of ensemble learning methods in traditional software engineering tasks—such as defect prediction, cost estimation, and quality assessment—this work first systematically introduces ensemble learning into the vulnerability detection of LLMs. We conduct a comprehensive empirical analysis of multiple ensemble strategies across different model architectures, task scenarios, and vulnerability types to assess the effectiveness and feasibility of ensemble learning in improving the accuracy and stability of LLM-based vulnerability detection.
\section{Conclusion}

This paper presents the first systematic study on the synergy between ensemble learning and LLMs for vulnerability detection, analyzing three common ensemble methods across five LLMs and three datasets. Our findings reveal that ensemble learning can significantly improve the performance of LLMs in vulnerability detection, with at least one ensemble method outperforming the baseline LLMs under various conditions. Based on our findings, we recommend using Boosting as the optimal choice when dealing with imbalanced data distributions, as Bagging and Stacking algorithms tend to underperform in such scenarios. The proposed DGS method, inspired by MoE, addresses Stacking's limitations with imbalanced data and multi-class tasks. Furthermore, based on our findings about the strengths of different ensemble learning methods and LLMs in detecting various types of vulnerabilities, we recommend further research into combining ensemble learning with MoE techniques to fully exploit their potential in this field.

\bibliographystyle{ACM-Reference-Format}
\bibliography{sample-base}


\end{document}